\documentclass[letterpaper,longbibliography,twocolumn,showpacs,floatfix,groupaddress,twocolumns]{revtex4-1}
\usepackage{enumerate}
\usepackage{float}
\usepackage{bm}
\usepackage[usenames]{color}
\usepackage{multirow}
\usepackage{amssymb}
\usepackage{amsbsy}
\usepackage{mathtools}
\usepackage{amsmath}
\usepackage{placeins}
\usepackage{bbold}
\usepackage{braket}
\usepackage{blindtext}
\usepackage[colorlinks,linkcolor=blue,citecolor=blue,urlcolor=blue]{hyperref}
\usepackage{siunitx}
\makeatletter

\newcommand{\Tr}[1]{\text{Tr}\left\{#1\right\}}

\begin{document}
\title{Typical perturbation theory: conditions, accuracy and comparison with a mesoscopic case}

\author{Mats H. Lamann}
\email{mlamann@uos.de}
\affiliation{Department of Physics, University of Osnabr\"uck, D-49069 
Osnabr\"uck, Germany}

\author{Jochen Gemmer}
\email{jgemmer@uos.de}
\affiliation{Department of Physics, University of Osnabr\"uck, D-49069 
Osnabr\"uck, Germany}

\begin{abstract}
The perturbation theory based on typicality introduced in Ref.\ \cite{FirstPaper} and further refined in \mbox{Refs.\ \cite{SpinGitter,LongPaper}} provides a powerful tool since it is intended to be applicable to a wide range of scenarios while relying only on a few parameters. Even though the authors present various examples to demonstrate the effectiveness of the theory, the conditions used in its derivation are often not thoroughly checked. It is argued that this is justified (without analytical reasoning) by the robustness of the theory. In the paper at hand, said perturbation theory is tested on three spin-based models. The following criteria are taken into focus: the fulfillment of the conditions, the accuracy of the predicted dynamics and the relevance of the results with respect to a mesoscopic case.

\end{abstract}

\maketitle

\section{Introduction}\label{Section1}
Typicality and randomness are fundamental concepts in modern quantum mechanics \cite{GemmerBook,PhysRevE.97.062129,Deutsch2018,DAlessio2016}.
A good example for randomness in quantum mechanics is the application of random matrix theory (RMT) in the eigenstate thermalization hypothesis (ETH) \cite{Deutsch,Srednicki,Rigol2008}, a concept of thermalization in isolated quantum systems.\\
Further, typicality is an important mechanism in quantum mechanics \cite{Balz2018,REIMANN2020121840,HeitmannRichterSchubertSteinigeweg+2020+421+432}.
The perturbation theory from Ref.\ \cite{FirstPaper} can be understood from the combination of both concepts: It is assumed that a perturbation $\hat{V}$ resulting in a perturbed Hamiltonian 
\begin{align}
\hat{H}=\hat{H}_0+\lambda\hat{V} , \label{eq:pert}
\end{align}
where $\hat{H}_0$ is the unperturbed Hamiltonian and $\lambda$ is the perturbation strength, can be considered as a random matrix (with the exception of some physically necessary properties). Therefore, one does not consider the perturbation for the individual case, but a whole ensemble of such perturbations.
In this way, a fairly general perturbation theory can be established, which depends only on a few para\-meters.
The remainder of the paper will be structured as follows:\\
First, in Section \ref{Section2}, the main-results and the respective conditions of the perturbation theory are summarized.
Then, in the Sections \ref{crossladder}-\ref{lattice}, the perturbation theory is examined using three different models. Within each Section, the model is first explained, the conditions of the theory are examined and then the results of the perturbation theory are compared with the numerical data. Finally, the results of each Section are summarized.\\
In Section \ref{mesos}, some results are compared with a mesoscopic case.
Section \ref{results} condenses all the results into a conclusion.

\section{Perturbation theory and conditions}\label{Section2}
According to the perturbation theory introduced in \mbox{Ref.\ \cite{FirstPaper}} (for the remainder of the paper, the phrase ``perturbation theory'' always refers to this kind of perturbation theory), the dynamics of a system perturbed as in Eq.\ \eqref{eq:pert} can be described at small $\lambda$ by the equation 
\begin{align}\label{eq:pert_g}
    \braket{\hat{A}}_{\Tilde{\rho}(t)} &= \braket{\hat{A}}_{\bar{\Tilde{\rho}}}+|g(t)|^2\left[\braket{\hat{A}}_{{\rho}(t)}-\braket{\hat{A}}_{\bar{\Tilde{\rho}}}\right] ,
\end{align}
where $\braket{\hat{A}}_{{\rho}(t)}$ is the unperturbed dynamics and $\braket{\hat{A}}_{\bar{\Tilde{\rho}}}$ is the long time value of the perturbed dynamics. It should be emphasized that $g(t)$ depends only on $\hat{V}$ and on the mean level spacing $\epsilon$ of the system but is independent of the observable $\hat{A}$.
There are various approximations for $g(t)$, which are derived under various assumptions. However, all approximations are based on the same approach, such that there are some fundamental conditions, which we list here:
\begin{enumerate}[i)]
    \item The density of states (DOS) in the non-perturbed system should be approximately constant if the level population of the initial state is not negligible, which implies that the local density of states (LDOS) is vanishingly small if the DOS is not constant. This is equivalent to an nearly constant mean level spacing $\epsilon$ for this relevant part.
    \item The perturbation should not change any thermodynamic quantities. This implies that the DOS should not be changed significantly either, because of its connection to entropy.
    \item The perturbation should be strong enough, such that the eigenstates of the unperturbed system $\ket{\nu}_0$ and those of the perturbed system $\ket{m}$ are well mixed. 
    \item The perturbation $\hat{V}$ should be a (pseudo) random matrix in the non-perturbed eigenbasis with the following properties
    \begin{align}
        \overline{\prescript{}{0}{\braket{\mu|\hat{V}|\nu}_0}}&=0\\
        \overline{|\prescript{}{0}{\braket{\mu|\hat{V}|\nu}_0}|^2}&=\sigma^2(|E_\mu-E_\nu|)
    \end{align}
    where the bar denotes the mean over an ensemble of perturbations and $\ket{\nu}_0$ is an eigenstate of the unperturbed Hamiltonian. The variance $\sigma^2$ is a smooth function of its arguments and will be called perturbation-profile within this paper.
\end{enumerate}
While the conditions ii) and iii) depend on $\lambda$, conditions i) and iv) do not.
Within this paper we always chose a large range of perturbation strengths $\lambda$, such that we consider condition iii) as fulfilled.\\
For small $\lambda$, 
\begin{align}\label{eq:g1}
    g_1(t)&=e^{-\Gamma \frac{|t|}{2}}\\
    &\text{with }\nonumber\\
    \Gamma &= 2\pi \lambda^2 \sigma^2(0)/\epsilon
\end{align}
is an appropriate approximation. On the other side, there is also an approximation for large $\lambda$. Note that this means, that $\lambda$ is large, but not so large that condition ii) is violated. This approximation takes the following form
\begin{align}\label{eq:g2}
    g_2(t)&=\frac{2J_1(\gamma t)}{\gamma t}\\
    &\text{with }\nonumber\\
    \gamma&=\lambda\sqrt{\frac{8\Delta_v\sigma^2(0)}{\epsilon}}\\
    \Delta_v&=\frac{1}{\sigma^2(0)}\int\limits_0^{\infty} \sigma^2(\omega)d\omega ,
\end{align}
where $J_1$ is the first kind Bessel function of order 1. The crossover between these approximations should happen at
\begin{align} \label{eq:lambda_c}
    \lambda_c&:=\sqrt{\frac{2\Delta_v\epsilon}{\pi^2\sigma^2(0)}}.
\end{align}

In a continuation of the theory \cite{LongPaper}, a third approximation
\begin{align}\label{eq:g3}
    g_3(t)&=\frac{\left(\gamma_+-\frac{\Gamma}{2}\right)e^{-\gamma_-|t|}+\left(\gamma_--\frac{\Gamma}{2}\right)e^{-\gamma_+|t|}-\Gamma e^{-\gamma_0|t|}}{2(\gamma_0-\Gamma)}\\
    &\text{with }\nonumber\\
    \gamma_{-,0,+}&=\frac{2\Delta_v}{\pi}\left[1\pm\sqrt{1-\frac{\pi \Gamma}{2\Delta_v}}\right]
\end{align}
was also determined, which, however, is analytically based on the fact that the profile of the perturbation is Lorentz-shaped
\begin{align}
    \sigma^2(\omega)&=\frac{\sigma^2(0)}{1+\left(\frac{\omega\pi}{2\Delta_v}\right)^2},
\end{align}
which is an additional condition for $g_3$, but is not necessary for the validity of $g_1$ and $g_2$.
It was shown in \mbox{Ref.\ \cite{LongPaper}} that $g_3$ is also a good approximation for some other profile shapes, but this is only a numerical check of some cases and might not be generally valid.\\
This approximation is also constructed for weak perturbations, but it is valid for larger ranges of $\lambda$ than $g_1$. For very weak perturbations $g_1$ and $g_3$ are identical. However, $g_2$ and $g_3$ differ even in the limit of strong and weak perturbations, as $g_3$ ($g_2$) is meant to hold for small (large) perturbation strengths $\lambda$.\\
There are other approximations for $g$, but these have more parameters than $g_l, \;l=\{1,2,3\}$. Therefore, we focus here on the approximations mentioned above, since a maximum of two parameters enables a more simple investigation. Moreover, these parameters are uniquely determined by $\Delta_v$ and ${\sigma^2(0)}/{\epsilon}$, such that if those properties of the perturbation are known, the different approximations can be compared with each other.

\section{spin ladder with cross-perturbation}\label{crossladder}
\begin{figure}[t]
    \centering
    \includegraphics[width=\linewidth]{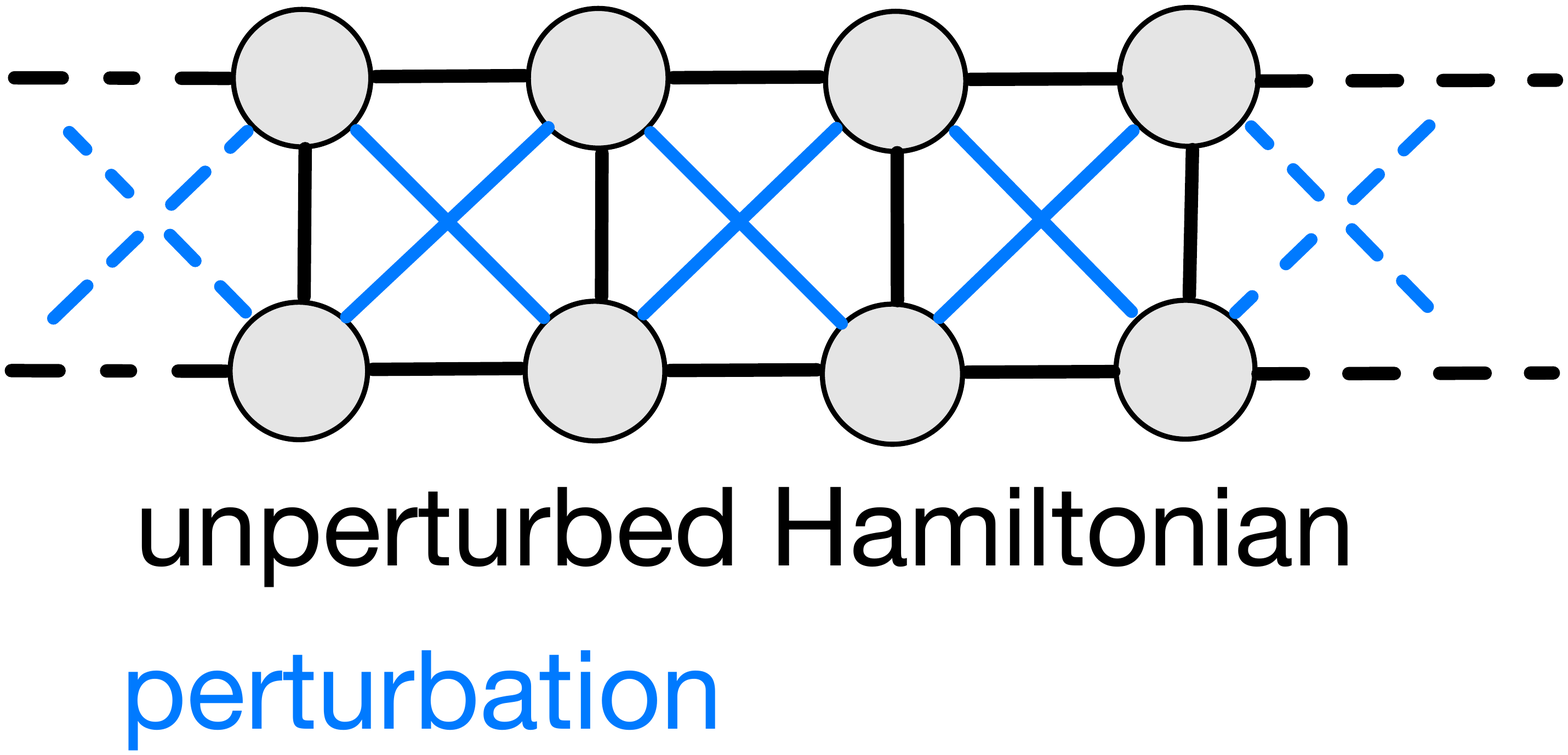}
    \caption{Sketch of the model from Section \ref{crossladder}:\\ Grey circles symbolize the spin sites, the black lines represent the Heisenberg interaction of the unperturbed Hamiltonian, the blue lines represent the perturbation. The dashed lines indicate the periodic boundaries of this system.}
    \label{fig:robinssystem_skizze}
\end{figure}
Our first system is a spin ladder with cross perturbations added (see Fig.\ \ref{fig:robinssystem_skizze}). The Hamiltonian $\hat{H}_0$ and the perturbation $\hat{V}$ take the following form
\begin{align}
    \hat{H}_0&=\hat{h}+\sum\limits_{j=1}^2\sum\limits_{l=1}^L \Vec{\hat{S}}_{l,j}\Vec{\hat{S}}_{l+1,j}+\sum\limits_{l=1}^L\Vec{\hat{S}}_{l,1}\Vec{\hat{S}}_{l,2}\label{eq:H_RS}\\
    \hat{h}&=-0.16\cdot\hat{S}^z_{1,1}+0.2\cdot\hat{S}^z_{4,2}+0.1\cdot\hat{S}^z_{5,2}\label{eq:h}\\
    \hat{V}&=\sum\limits_{l=1}^L\hat{S}^z_{l,1}\hat{S}^z_{l+1,2}+\hat{S}^z_{l,2}\hat{S}^z_{l+1,1}\label{eq:V_RS}
\end{align}
with periodic boundary conditions $\Vec{\hat{S}}_{L+1,k}\widehat{=}\Vec{\hat{S}}_{1,k}$. Here $\Vec{\hat{S}}_{l,j}$ are the standard spin-$1/2$ operators acting on site $l$ of the $j$-th leg.
In our research, we focus on the magnetization-subspace with vanishing total magnetization in $z$-direction. This, in combination with $\hat{h}$, ensures that there are no symmetries within this subspace. This is necessary since the perturbation theory relies on the independence of matrix elements, and symmetries cause structured sparseness (on the other hand, unstructured sparseness is not in conflict with the theory). In this paper we always investigate the smallest possible subspace within each system.\\
We investigate the dynamics of the observable 
\begin{align}
    \hat{S}_q&=\sum\limits_{l=1}^L \cos{\left(\frac{2\pi}{L}\cdot l\right)}\left(\hat{S}^z_{l,1}+\hat{S}^z_{l,2}\right) ,
\end{align}
which corresponds to the slowest mode of magnetization in the $z$-direction of a rung and how they behave under the perturbation from Eq.\ \eqref{eq:pert}.
\begin{figure}[t]
  \includegraphics[width=\linewidth]{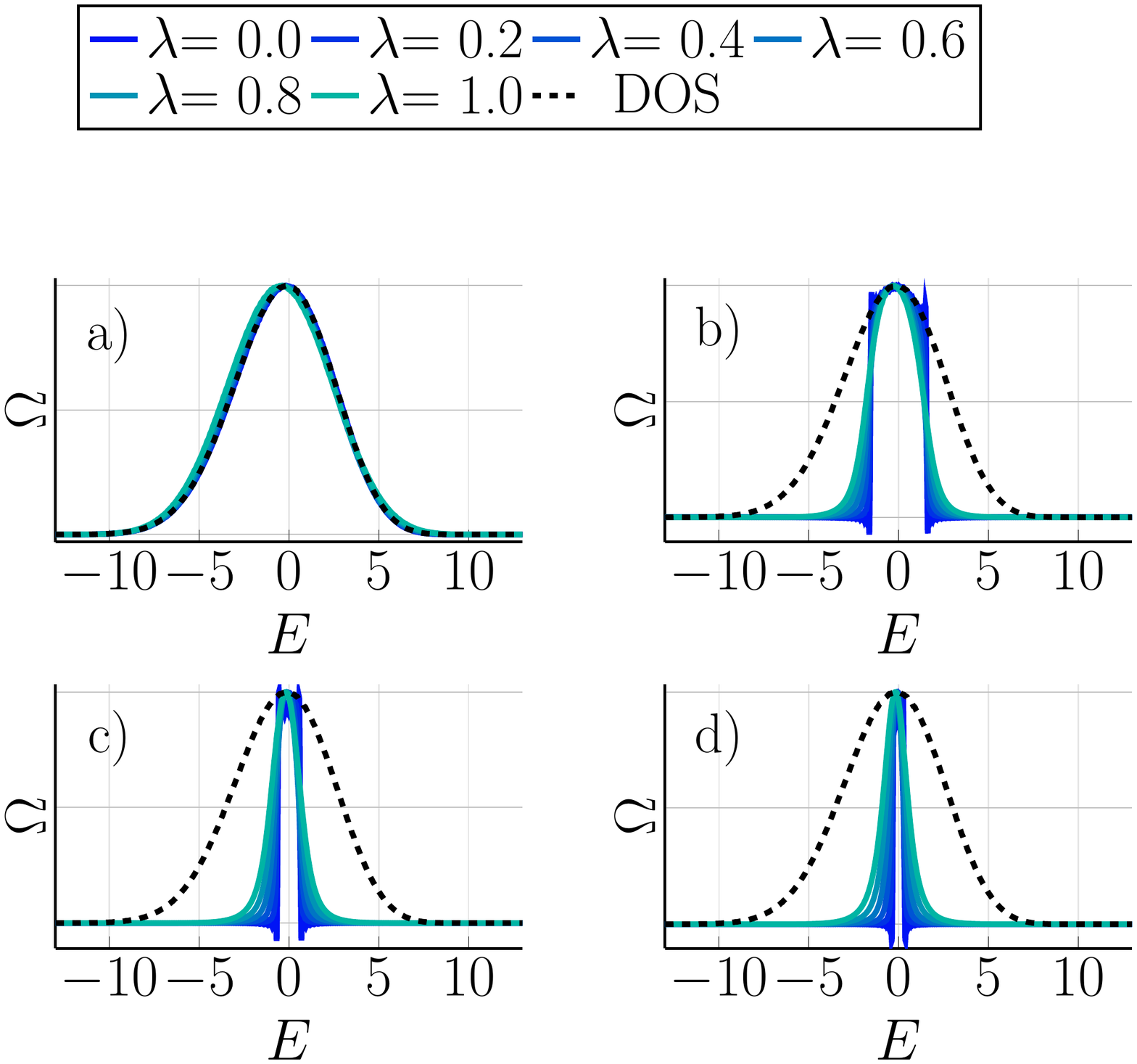}
  \caption{LDOS for various perturbations strength $\lambda$ and energy windows in the system of Section \ref{crossladder}.\\ a) full spectrum, b) $\Delta E=\frac{\pi}{2}$, c) $\Delta E=\frac{\pi}{5}$ and d) $\Delta E=\frac{\pi}{10}$.\\ (Scaled for better comparability)}
  \label{fig:RSDOS} 
\end{figure}
This system has already been studied in Ref.\ \cite{HevelingKnipschildGemmer+2020+475+481}, also in comparison with the perturbation theory. Therein it was stated that the perturbation theory is not suitable for some dynamics within this system. However, the conditions of the perturbation theory were not examined more closely.\\
In particular, in Ref.\ \cite{HevelingKnipschildGemmer+2020+475+481} the LDOS is also populated where the DOS is not constant. Therefore, we choose to investigate this model in different energy windows. If the deviation of the perturbation theory found in Ref.\ \cite{HevelingKnipschildGemmer+2020+475+481} is related to the violation of \mbox{condition i),} it is expected that the perturbation theory for sufficiently small windows can also describe this model.\\
In this system, we consider an initial state
\begin{align}
    \hat{\rho} &\propto \mathcal{P}_{E,\Delta E}\left(\hat{S}_q-\kappa\hat{1}\right)\mathcal{P}_{E,\Delta E}\\
    &\text{with}\nonumber\\
    \mathcal{P}_{E, \Delta E}  &=   \sum\limits_{|E^0_{\nu}-E|<\Delta E} \ket{\nu}_0 \prescript{}{0}{\bra{\nu}},
\end{align}
where $\mathcal{P}_{E,\Delta E}$ is an projector to an energy window centered around $E$ with a width of $2\Delta E$ (in the unperturbed Hamiltonian), $\kappa$ is the smallest eigenvalue of $\hat{S}_q$ and $\hat{1}$ is the identity. The sum runs over all energy eigenstates $\ket{\nu}_0$ of the unperturbed Hamiltonian with $|E^0_\nu|\leq \Delta E$.\\\
\begin{figure}
    \includegraphics[width=\linewidth]{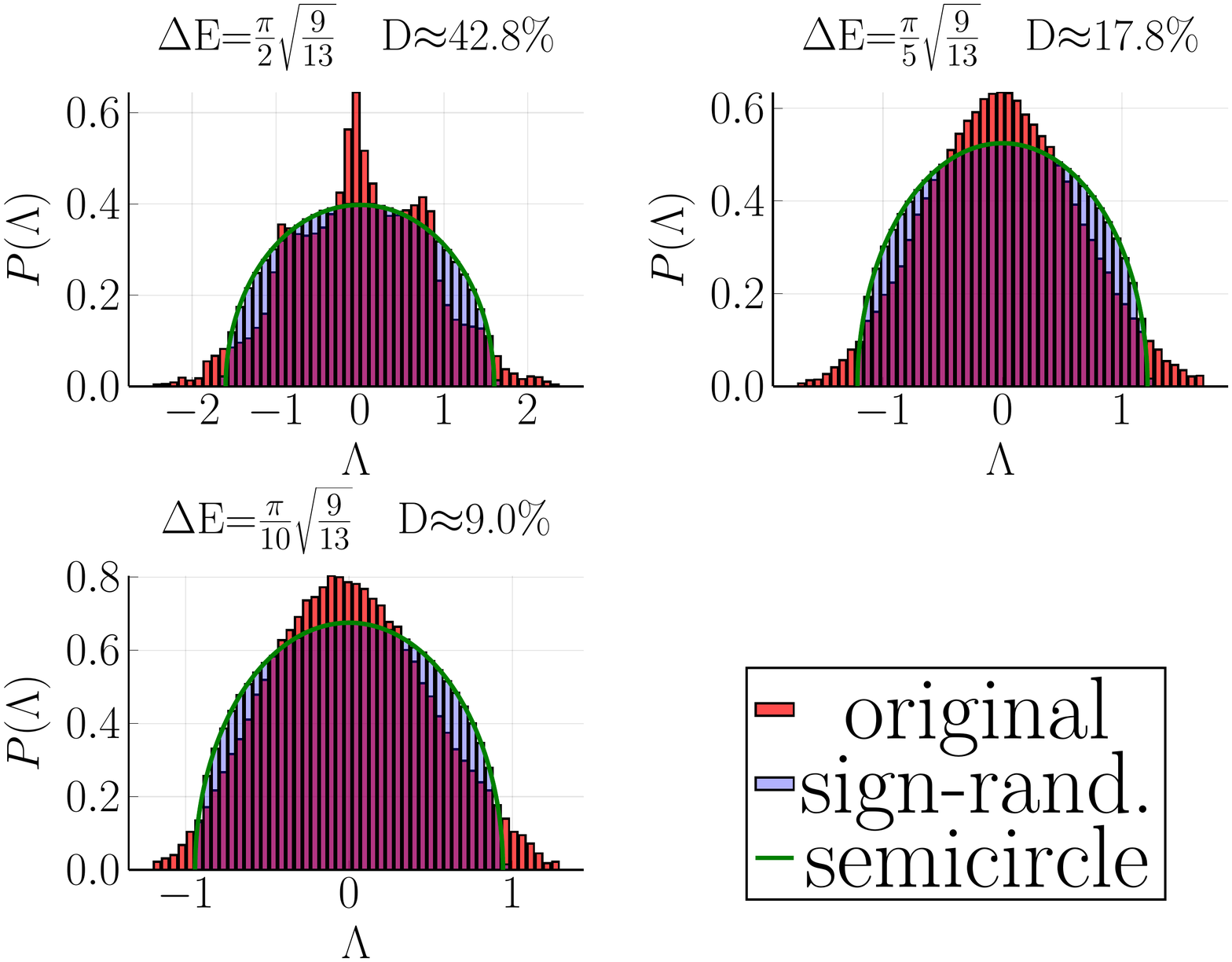}
    \caption{Spectra of the (trace free) perturbation $\hat{V}_{\Delta E}$ and the sign-randomized version $\Tilde{V}$ for different energy windows $\Delta E$, for $L=9$. In addition the results are compared with the Wigner-semicircle law, which gives the distribution for a random matrix. $D$ indicates what percentage of the total system is in the chosen energy window.}
    \label{fig:SignRandCross}
\end{figure}
The dynamics are given by
\begin{align}
    \braket{\hat{S}_q(t)}&:=\Tr{\hat{\rho}\hat{S}_q(t)}\\
    \text{with}\nonumber\\
    \hat{S}_q(t)&=e^{i\hat{H}t}\hat{S}_qe^{-i\hat{H}t},
\end{align}
where we set $\hbar=1$. This corresponds to the autocorrelation function (with some constant shift) in the limit of $\lambda\rightarrow 0$ or $\mathcal{P}_{E,\Delta E}\rightarrow\hat{1}$:
\begin{align}
    C_{\hat{S}_q}(t)&=\frac{\Tr{\hat{S}_q(t)\mathcal{P}_{E,\Delta E}\hat{S}_q\mathcal{P}_{E,\Delta E}}}{\Tr{\mathcal{P}_{E,\Delta E}}}\\
    &\propto\braket{\hat{S}_q(t)}+C.
\end{align}
With the help of $\mathcal{P}_{E,\Delta E}$, the dependence of the perturbation theory on condition i) can be investigated. The projector $\mathcal{P}_{E,\Delta E}$ does not create a sharp energy window in $\hat{H}$, but nevertheless the energy range should be reasonably sharp (see Fig.\ \ref{fig:RSDOS}) for weak perturbations.
\subsection{Conditions}
\begin{figure}[t]
  \includegraphics[width=\linewidth]{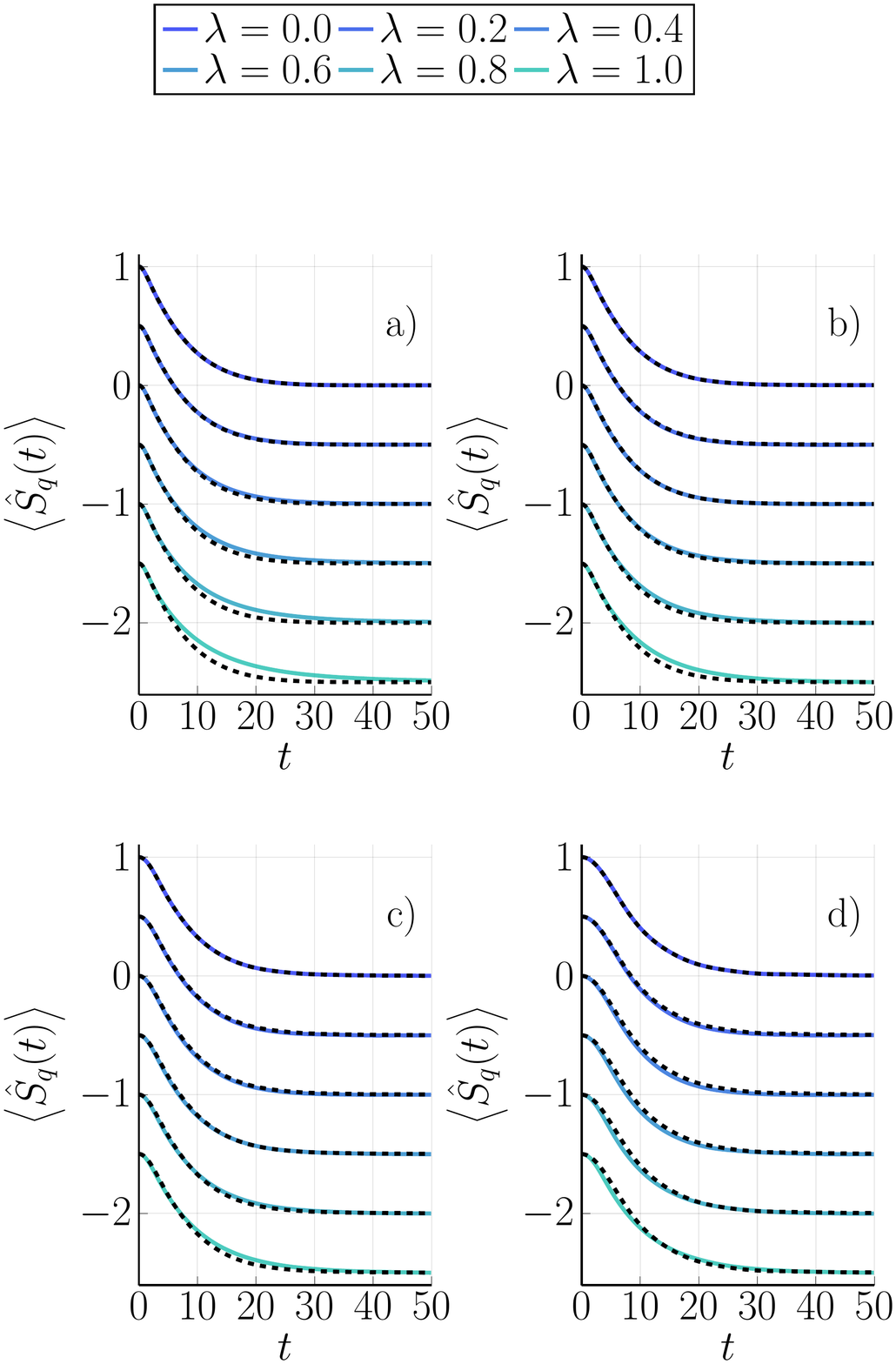}
  \caption{Dynamics for various perturbations strength $\lambda$ and energy windows in the system of Section \ref{crossladder}. The dashed lines denote the unperturbed dynamics within the corresponding energy window. The curves are normalized and shifted by $0.25$.\\ a) full spectrum, b) $\Delta E=\frac{\pi}{2}$, c) $\Delta E=\frac{\pi}{5}$ and d) $\Delta E=\frac{\pi}{10}$.}
  \label{fig:Dyn_RS} 
\end{figure}
To investigate conditions i) and ii), the LDOS (DOS) $\Omega$ of the initial state (the system) must be determined for various perturbation strengths and energy window sizes.
Since the system with $L=13$ ($\mathcal{D}=10400600$) is much too large for exact diagonalisation (ED), the DOS and also the LDOS
\begin{align}
    \Omega \propto \sum\limits_{|E-E_m|<\delta E} \langle m|\hat{\rho}|m \rangle,
\end{align}
where $\delta E$ is a macroscopically small but microscopically large energy window, can not be determined exactly here. With the help of typicality methods we can estimate the LDOS with high precision \cite{HeitmannRichterSchubertSteinigeweg+2020+421+432}.\\
The system is examined within different energy windows (of the unperturbed Hamiltonian). The size of these windows compared to the whole spectrum can be seen in Fig.\ \ref{fig:RSDOS}. The windows are defined with the central energy $E=0$ and the widths $\Delta E = \left\{\infty, \frac{\pi}{2}, \frac{\pi}{5}, \frac{\pi}{10}\right\}$.\\\
\begin{figure}[t] 
  \includegraphics[width=\linewidth]{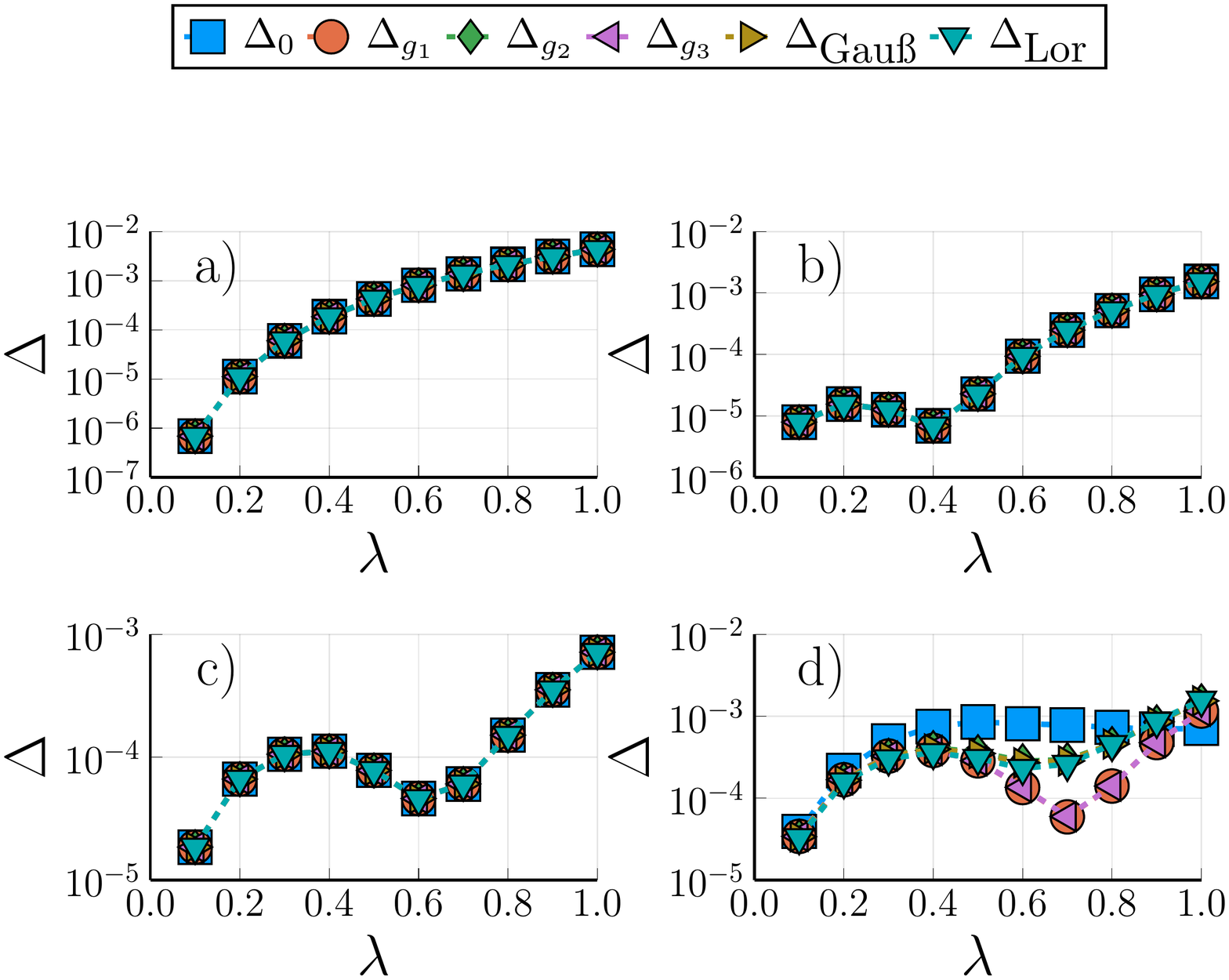}
  \caption{Deviation between the prediction of the perturbation theory and perturbed dynamics as well as the deviation between the perturbed and unperturbed dynamics for various perturbations strength $\lambda$ and energy windows in the system of Section \ref{crossladder}. \\a) full spectrum, b) $\Delta E=\frac{\pi}{2}$, c) $\Delta E=\frac{\pi}{5}$ and d) $\Delta E=\frac{\pi}{10}$.}
  \label{fig:Delta_RS} 
\end{figure}
The projector $\mathcal{P}_{E,\Delta E}$ onto the respective energy window is achieved with a method from \mbox{Ref.\ \cite{PhysRevLett.128.180601}}, which works without ED.
As can be seen in Fig.\ \ref{fig:RSDOS}, the smaller the energy window, the better condition i) is satisfied.
Moreover, it can be noticed from a) in Fig.\ \ref{fig:RSDOS} that the DOS of the full spectrum is never strongly altered by the perturbation. Therefore, condition ii) is also fulfilled.\\
The deviations of the LDOS from the hard energy cut can be explained by the fact that the energy window is chosen in $\hat{H}_0$, but the LDOS is studied in the perturbed Hamiltonian $\hat{H}$.\\
Condition iv) is checked using the sign-randomization method \cite{PhysRevE.102.042127}.
In this method, the spectrum of a operator $\hat{V}$ is compared with a sign-randomized version this operator
\begin{align}
    \Tilde{V}_{mn}&=
    \begin{cases}
            V_{mn}      & 50\%\\
            -{V}_{mn}   & 50\%,
    \end{cases}
\end{align}
where the elements are randomly multiplied by $1$ or $-1$ (while keeping the operator hermitian). This destroys possible correlations between the elements.\\
Since a non-zero trace always changes under sign-randomization, we use this method on a trace free version $\hat{V}_{\Delta E}$ of the perturbation in a given energy window:
\begin{align}
    \hat{V}_{\Delta E} &=   \mathcal{P}_{E,\Delta E}\hat{V}\mathcal{P}_{E,\Delta E}-\frac{\Tr{\mathcal{P}_{E,\Delta E}\hat{V}\mathcal{P}_{E,\Delta E}}}{\Tr{\mathcal{P}_{E,\Delta E}}}\cdot\mathcal{P}_{E,\Delta E}
\end{align}
Since the comparison of the spectra needs ED, this method is limited to sizes up to $L=9$.
To take the scaling of the system into account, the windows are modified: $\Delta E = \left\{\frac{\pi}{2}, \frac{\pi}{5}, \frac{\pi}{10}\right\}\cdot\sqrt{\frac{9}{13}}$.
This scaling is justified by the reason, that the energy standard deviation of such a system scales with the root of the system size.\\
The spectra are depicted in Fig.\ \ref{fig:SignRandCross}. It is easy to see that there are deviations between the spectra even in small windows, such that condition iv) is not fulfilled in these cases.
Even though the correlations are only checked for smaller system sizes, there are indications that even for larger systems the correlations do not fully vanish even after relatively long times (compared to the relaxation time) \cite{PhysRevLett.128.180601}, which corresponds to small energy windows.
\begin{figure}[t]
  \includegraphics[width=\linewidth]{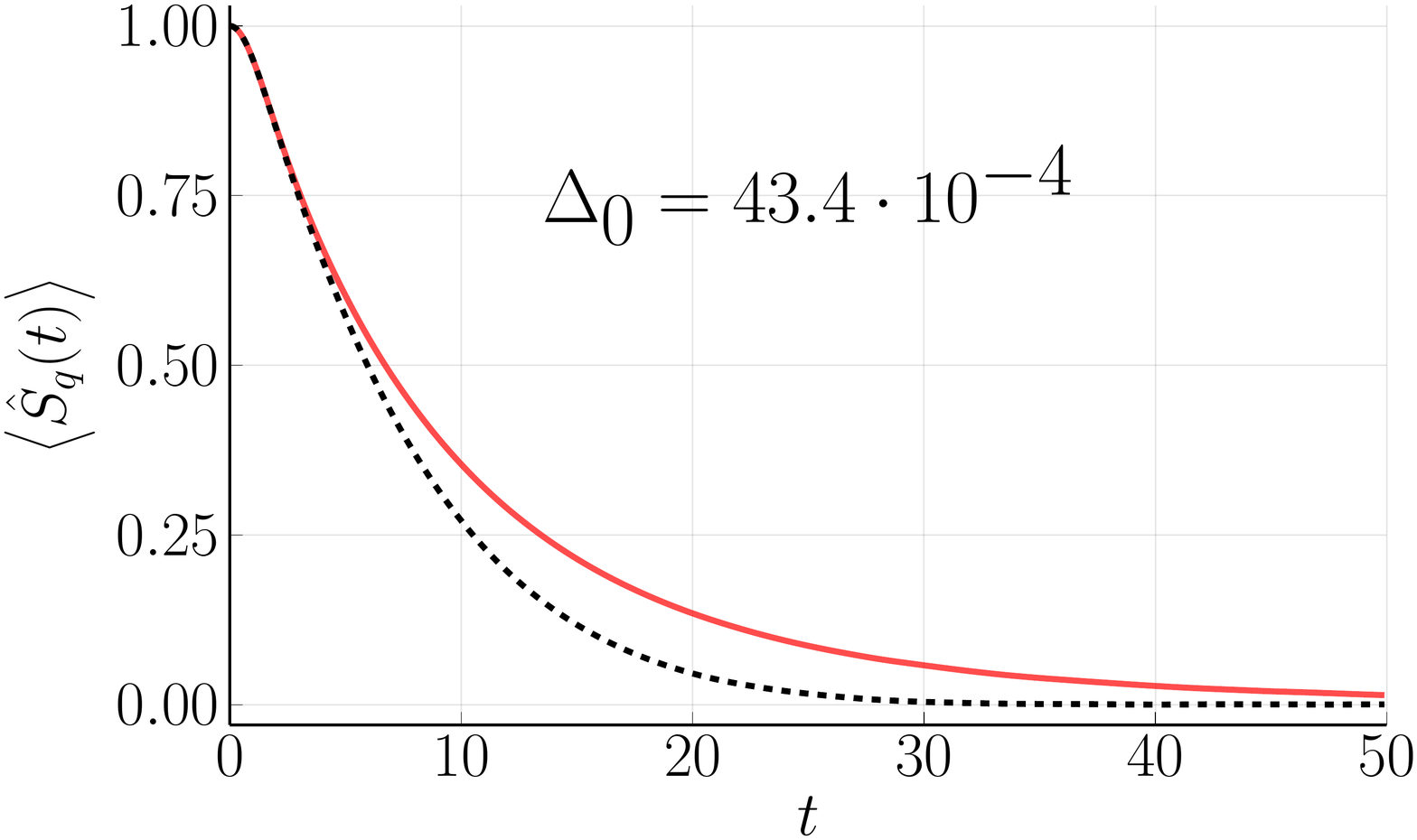}
  \caption{Example of the difference between perturbed (solid) and unperturbed (dashed) dynamics. ($\Delta E=\infty$, $\lambda=1.0$)}
  \label{fig:Example_RS} 
\end{figure}
\subsection{Comparison}
To quantify the accuracy of the perturbation theory, the deviation 
\begin{align}
    \Delta_{g_l}(\lambda) &= \frac{1}{\tau \braket{\hat{S}_q(0)}_\text{num.}^2}\int\limits_0^{\tau} \left|\braket{\hat{S}_q(t)}_{g_l} - \braket{\hat{S}_q(t)}_\text{num.}\right|^2 \text{d}t \label{eq:Delta}
\end{align}
is observed. Here $\braket{\hat{S}_q(t)}_{g_l}$ are the dynamics predicted by the perturbation theory [cf.\ Eq.\ \eqref{eq:pert_g}] for the approximation ${g_l}$, $\braket{\hat{S}_q(t)}_\text{num.}$ are the directly numerically determined dynamics %%(via chebycheff method)
and $\tau$ is the relaxation time of the perturbed dynamics. Although there are many various definitions of relaxation times, we will choose a rather plain one here. We here define the relaxation time $\tau$ as the time when \mbox{$\Tilde{C}(t)=[C(t)-C(t\rightarrow\infty)]/[C(0)-C(t\rightarrow\infty)]$} has decayed to $\Tilde{C}(t)<0.01$ and stays below this threshold afterward. \\
% Note that the relaxation time of the perturbed system may differ from the relaxation time of the unperturbed system. In this case, the longer time is always chosen as the limit of integration.\\
In addition, the deviation
\begin{align}
    \Delta_0(\lambda) &= \frac{1}{\tau \braket{\hat{S}_q(0)}_\text{num.}^2}\int\limits_0^{\tau} \left|\braket{\hat{S}_q(t)}_0 - \braket{\hat{S}_q(t)}_\text{num.}\right|^2 \text{d}t, \label{eq:Delta_0}
\end{align}
where $\braket{\hat{S}_q(t)}_0$ denotes the unperturbed dynamics, is also considered and compared with the previous deviations.\\
Even though strict testing of the perturbation theory would require the parameters for the approximations $g_l$ to be determined from the perturbation operator, those parameters ($\sigma^2 (0)$,$\Delta_v$) are treated as fitting parameters since a direct determination is not possible due to the large dimension of the system.\\A suitable  description of the dynamics by means of fitting is a necessary property for the correctness of the theory but not a sufficient one.\\
The fitting is optimized in such a way, that it minimizes the total deviation
\begin{align}
    \Delta_{tot}&=\sum_{i=0}^N\frac{\Delta_{g_l}(\lambda_i)}{\lambda_i} , 
\end{align}
where the sum runs over all investigated perturbation strengths $\lambda_i$ for each $g_l$ and energy window/filter. The weighting is reasoned by the fact, that the perturbation theory are not meant to hold for too strong perturbations.\\
Even though one could choose another weighting (e.g. $\lambda_i^{-2}$), which increases for weaker perturbation, the overall results do not depend strongly on this choice.\\
Fig.\ \ref{fig:Dyn_RS} shows the dynamics of this system. The initial value was normalized to $1$. The dashed black lines indicate the unperturbed dynamics.
It can be seen that the perturbation affects the curves differently depending on the energy window. Thus, for large or no energy windows, the unperturbed dynamics decays always faster than the perturbed dynamics. For smaller windows, on the other hand, a faster relaxation is shown for small times by perturbed systems compared to the unperturbed case. \\
Since the perturbation theory can only accelerate the perturbed dynamics (cf.\ Section \ref{Section1}), the dynamics for large windows cannot be explained by the perturbation theory. 
The deviation $\Delta_{g_l}$ between the prediction by the perturbation theory and the numerically determined perturbed dynamics are displayed in \mbox{Fig.\ \ref{fig:Delta_RS}}.
For better illustration, Fig.\ \ref{fig:Example_RS} also shows a comparison of perturbed and unperturbed dynamics, which correspond to the largest specific deviation in this system ($\Delta_0=43.4\cdot 10^{-4}$).
It can be seen from Fig.\ \ref{fig:Delta_RS} that the perturbation theory is unsuitable for most energy windows. The deviations between the theory and the perturbed curves coincides with the difference between the unperturbed and the perturbed dynamics, which is the worst possible result, since the perturbation theory always includes the unperturbed dynamics as limit of $\sigma^2(0)\rightarrow 0$.\\
In addition to the perturbation theories from Ref.\ \cite{LongPaper}, alternative functions without a theoretical basis for $g$ are also tested:
\begin{align}
    g_{\textrm{Gau}}&=\exp{\left(-\alpha\cdot\left(\lambda t\right)^2\right)}\label{eq:gau}\\
    g_{\textrm{Lor}}&=\frac{1}{1+\alpha\cdot(\lambda t)^2}\label{eq:lor}
\end{align}
These Functions were chosen such that $g(0)=1$ and $g(\infty)=0$, which are necessary for Eq.\ \eqref{eq:pert_g} to satisfy at least the trivial properties of the perturbed dynamics. The variable $\alpha$ is a free fitting parameter.\\
Even for the smallest energy window, where the perturbation theory in general describes the change of the dynamics better than the unperturbed dynamics, the expectation for the different approximations $g_l$ is not fulfilled: Even though $g_1$ and $g_3$ are just approximations for very weak perturbation, they describe the strong perturbed dynamics better than $g_2$, which is meant for the case of large perturbation strengths $\lambda$. 
The fact that the perturbation theory generally yields good results in this window can also be explained outside the theory:\\
Since the windows have been chosen in the unperturbed system, any perturbations will enable higher frequencies.
It is not surprising that these newly available frequencies provide faster relaxation for small times, since the initial state chosen is far from the equilibrium state. This explanation is supported by the fact that the faster relaxation only occurs for small times.\\
The resulting dynamics for the arbitrarily chosen functions $g_{\textrm{Gau}}$ ($g_{\textrm{Lor}}$) exhibit the similar behavior as the results from the perturbation theory, as they only differ from the unperturbed dynamics in the smallest energy window.\\
In Fig.\ \ref{fig:Delta_RS} one can see, that the results from this function are close the results for $g_2$, while $g_{1,3}$ yield better results.\\
Even though the perturbation theory trivially get better results for small windows, since newly available frequencies accelerate the dynamics, we can see that other arbitrarily (which also only accelerate the dynamics) do not yield such good results as the known \mbox{approximations for $g$}.
\begin{figure}[t]
    \centering
    \includegraphics[width=\linewidth]{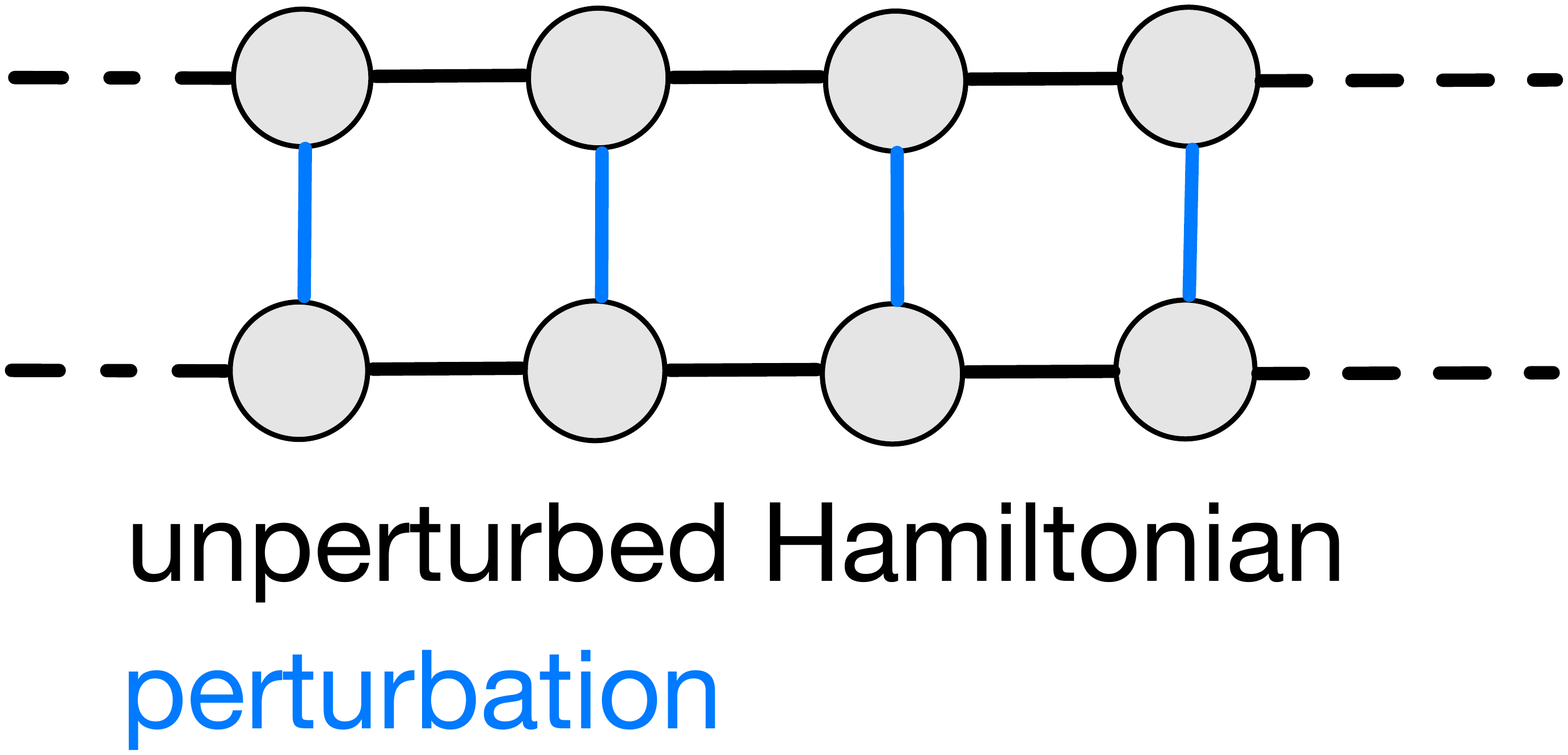}
    \caption{Sketch of the model from Section \ref{chainladder}:\\ Grey circles symbolize the spin sites, the black lines represent the Heisenberg interaction of the unperturbed Hamiltonian, the blue lines represent the perturbation. The dashed lines indicate the periodic boundaries of this system.}
    \label{fig:jonasleiter_skizze}
\end{figure}
\subsection{Conclusion}
For this system it could be shown that the assumption that the DOS is constant (condition i)) is not generally true. However, with the help of energy windows, it is possible to investigate dynamics in this system that do fulfill condition i).
Even in these very small energy windows, we were still able to prove correlations, so that condition iv) is never fulfilled.\\
The DOS itself remained for the most part unchanged under the examined perturbation strengths, so that we can consider dynamics that fulfill both \mbox{condition i) and ii)}.
The perturbation theory only gives good results, if the window size is very small. In this case, however, an improvement can be explained without using the perturbation theory.
\section{spin chain to spin ladder}\label{chainladder}
The Hamiltonian and the perturbation of the second model is given by
\begin{align}
    \hat{H}_0&=\hat{h}+\sum\limits_{j=1}^2\sum\limits_{l=1}^L \Vec{\hat{S}}_{l,j}\Vec{\hat{S}}_{l+1,j}\label{eq:H_JS}\\
    \hat{V}&=\sum\limits_{l=1}^L\Vec{\hat{S}}_{l,1}\Vec{\hat{S}}_{l,2}\label{eq:V_JS}
\end{align}
again with periodic boundary conditions and the symmetry breaking term $\hat{h}$ from Eq.\ \eqref{eq:h}. As one can see in \mbox{Fig.\ \ref{fig:jonasleiter_skizze}}, this perturbation connects two separated spin chains into a spin ladder.
Note that our model differs from the model in Ref.\ \cite{LongPaper} only by the symmetry breaking term $\hat{h}$.\\
The observable in this system is the spin current along the ladder legs
\begin{align}
    \hat{J}&=\sum\limits_{j=1}^2\sum\limits_{l=1}^L \hat{S}^x_{l,j}\hat{S}^y_{l+1,j}-\hat{S}^y_{l,j}\hat{S}^x_{l+1,j}.
\end{align}
The autocorrelation function
\begin{align}
    C_{\hat{J}}(t)&=\frac{\Tr{\hat{J}(t)\hat{J}}}{\mathcal{D}} ,
\end{align}
where $\mathcal{D}$ is the dimension of the Hilbertspace, is considered in these systems. 
The autocorrelation function corresponds to the dynamics of a density matrix
\begin{align}
    C_{\hat{J}}(t)&\propto\Tr{\hat{\rho}\hat{J}(t)}=:\braket{\hat{J}(t)}\\
    \hat{\rho}&\propto \hat{1}+\zeta\hat{J}\label{eq:rho_J}
\end{align}
with an sufficiently small $\zeta$. This has some similarities with the model in Section \ref{crossladder} for the unfiltered case, since both dynamics correspond to an autocorrelation function.
The projection into energy windows was omitted because in Ref.\ \cite{LongPaper} perturbation theory yielded good results even without filters.\\\
\begin{figure}[t]
    \includegraphics[width=\linewidth]{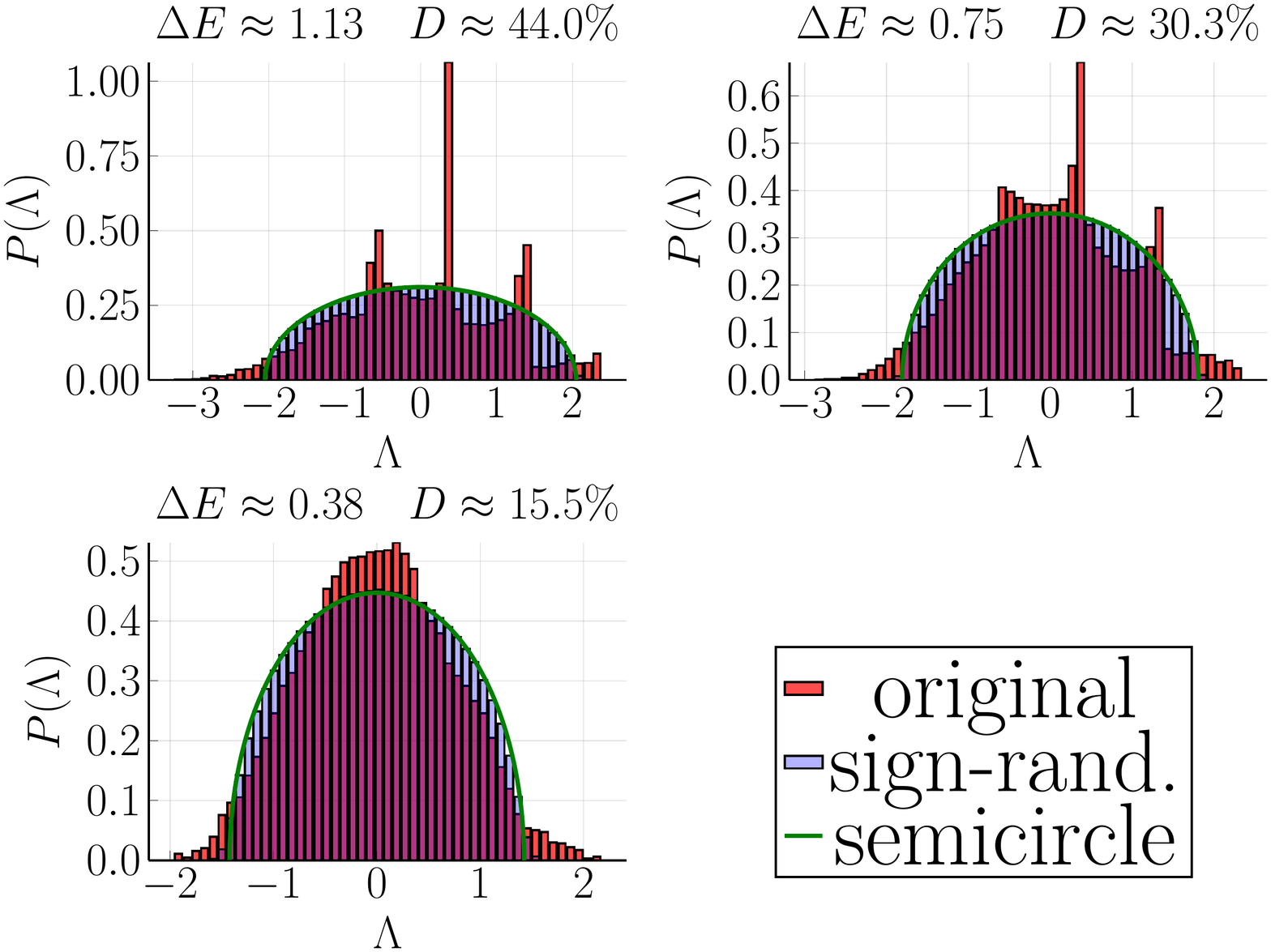}
    \caption{Spectra of the (trace free) perturbation $\hat{V}_{\Delta E}$ and the sign-randomized version $\Tilde{V}$ for different energy windows $\Delta E$, for $L=9$. In addition the results are compared with the Wigner-semicircle law, which gives the distribution for a random matrix. $D$ indicates what percentage of the total system is in the chosen energy window.}
    \label{fig:SignRandChain}
\end{figure}
The unperturbed system is close to integrable and transitions to a non-integrable system if perturbed.\\
The second model was studied in detail in Ref.\ \cite{JonasSpinLeiter} and compared with perturbation theory results in \mbox{Ref.\ \cite{LongPaper}}. There, perturbed dynamics could be well described by the perturbation theory, even though condition i) is trivially not satisfied [cf.\ Eq.\ \eqref{eq:rho_J}]. The parameters $\sigma^2(0)$ and $\Delta_v$ were not determined but treated as free fitting parameters, which is a necessary property not a sufficient one. \\
\subsection{Conditions}
\begin{figure}[t]
    \centering
    \includegraphics[width=\linewidth]{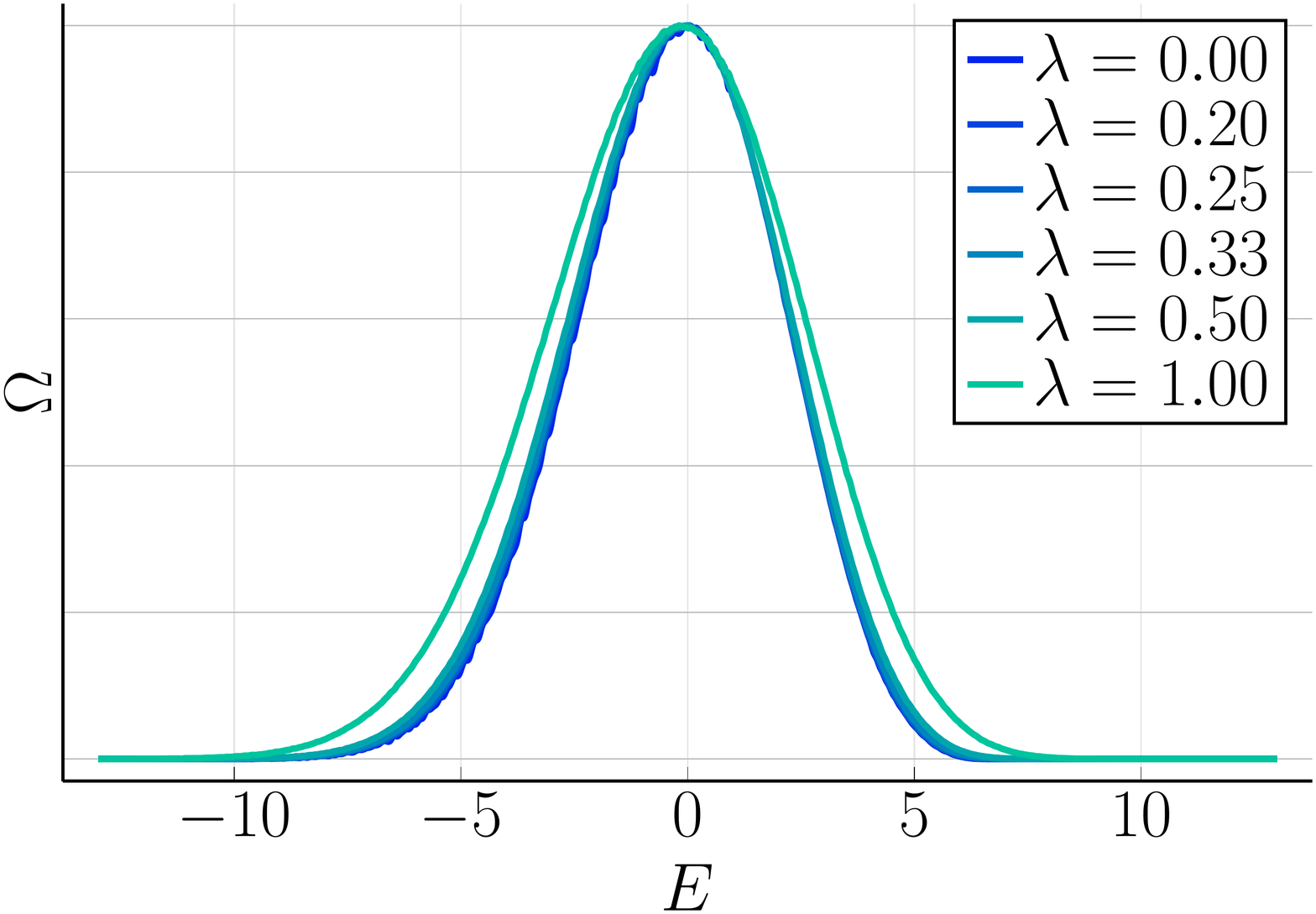}
    \caption{DOS at different perturbation strengths $\lambda$ from the spin ladder described in Section \ref{chainladder}. (Scaled for better comparability)}
    \label{fig:DOS_JS}
\end{figure}

Condition i) is trivially not fulfilled. This is easily seen from Eq.\ \eqref{eq:rho_J}, since the LDOS of this density matrix for small $\zeta$ is for no energy negligibly small (with exception when the DOS is negligibly small). \\
In Fig.\ \ref{fig:DOS_JS} it can be seen that the DOS does not change much at the different perturbation strengths $\lambda$ and a system size $L=13$, so condition ii) is satisfied up to $\lambda=1.0$.\\
For the condition iv) we have to limit ourselves to system sizes within the range of ED, so the sign-randomization method is applied to a system with $L=9$, which can be seen in Fig.\ \ref{fig:SignRandChain}. Thereby we chose as energy windows 
\begin{align}
    \Delta E=\left\{0.6,0.4,0.2\right\}\cdot\sqrt{\frac{\Tr{\hat{H_0}^2}}{\mathcal{D}}},
\end{align}
such that they are small compared to the standard deviation of the full spectrum of the unperturbed Hamiltonian.\\
It is easy to see that even in narrow energy windows the spectra differ greatly from each other. Since the dynamics of this system are not restricted to a small energy window, even the correlation in larger windows contribute to the dynamics.
In summary, conditions i) and iv) are not fulfilled in this system.
\subsection{Comparison}
\begin{figure}[t]
    \includegraphics[width=\linewidth]{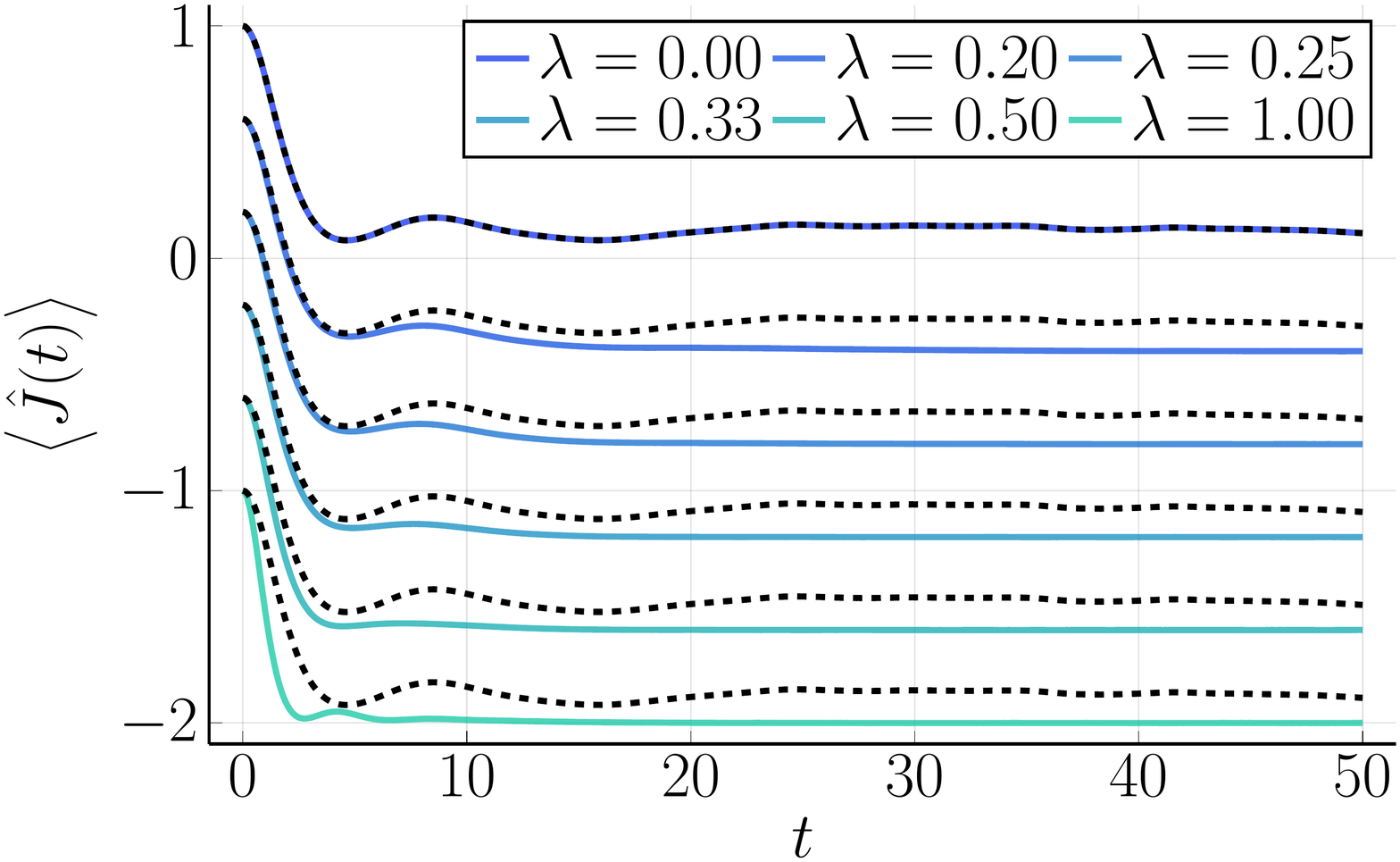} 
    \caption{Dynamics for various perturbations strength $\lambda$ in the system of Section \ref{chainladder}. The dashed lines denote the unperturbed dynamics. The curves are normalized and shifted by $0.4$.}
    \label{fig:Dyn_JS} 
\end{figure}

\begin{figure}[t]
    \includegraphics[width=\linewidth]{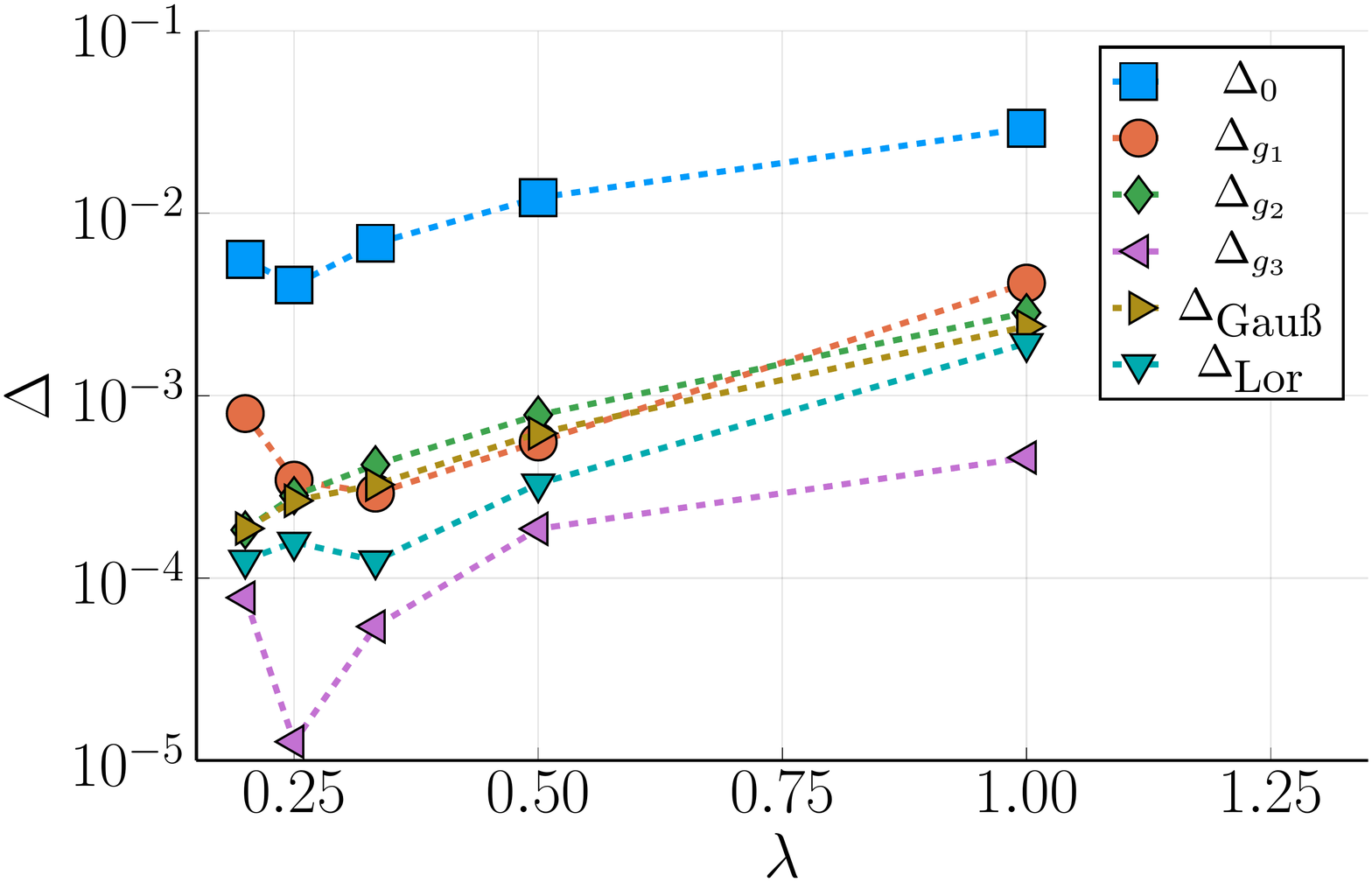} 
    \caption{Deviation between the prediction of the perturbation theory and perturbed dynamics as well as the deviation between the perturbed and unperturbed dynamics in the system of Section \ref{chainladder}.}
    \label{fig:Delta_JS} 
\end{figure}

Since the investigated dynamics are for system size beyond the range of ED ($L=13$), the parameters are again not be determined directly but treated as fitting parameters.
The dynamics of the system from Eq.\ \eqref{eq:H_JS} can be seen in Fig.\ \ref{fig:Dyn_JS}. The black lines show the unperturbed dynamics.
This makes it easy to see that the long time value changes considerably due to the perturbation. \\
For this reason, the deviations of the unperturbed dynamics to the perturbation theory is mainly determined by the long time value. We emphasize that the long time value of the perturbed dynamic is always a fitting parameter in the perturbation theory, since the theory provides no method for determining this value.\\
The deviations are shown in \mbox{Fig.\ \ref{fig:Delta_JS}}.
It can be seen that the perturbation theory is at least a better description compared to the unperturbed case.\\
Here it is easy to see that $g_3$ provides the best description of the perturbation. The approximation $g_1$ is even for weak perturbations worse than $g_2$, which contradicts the expectation.\\
However, functions chosen without a theoretical basis [Eq.\ \eqref{eq:gau} and \eqref{eq:lor}] results in better descriptions of the perturbed dynamic than $g_1$ and $g_2$.\\
Although the results of $g_3$ are the best, it should be mentioned that this approximation also has two free parameters, while the other function have only one, so that the range of possible functions is broader in the case of $g_3$.
\subsection{Conclusion}
Even though this system is listed in Ref.\ \cite{LongPaper} as an example in favor of the perturbation theory, already \mbox{condition i),} that the DOS is constant for the relevant part, is not fulfilled.
Condition ii) is given in all cases except $\lambda=1.00$, while condition iv) is never fulfilled. \\
In this system the perturbation theory shows a certain benefit, so the deviations between perturbation theory and perturbed dynamics are smaller than between unperturbed dynamics and perturbed dynamics, however, also relatively arbitrarily chosen functions show a partly even more accurate description. Therefore, it cannot be ruled out that the validity of the perturbation theory is accidental here and does not necessarily follow from the derivation in Refs.\ \cite{FirstPaper,LongPaper}. \\
Moreover, the parameters of the perturbation theory have been treated here as free fitting parameters. Although it is necessary for the validity of the perturbation theory that such a fitting yields good results, it is not sufficient, since the true parameters could still yield poor results.
It also appears that in this case the deviations are partly caused by the differences of the long-term values, which cannot be determined within the theory.
\section{spin lattice}\label{lattice}
\begin{figure}[t]
    \centering
    \includegraphics[width=\linewidth]{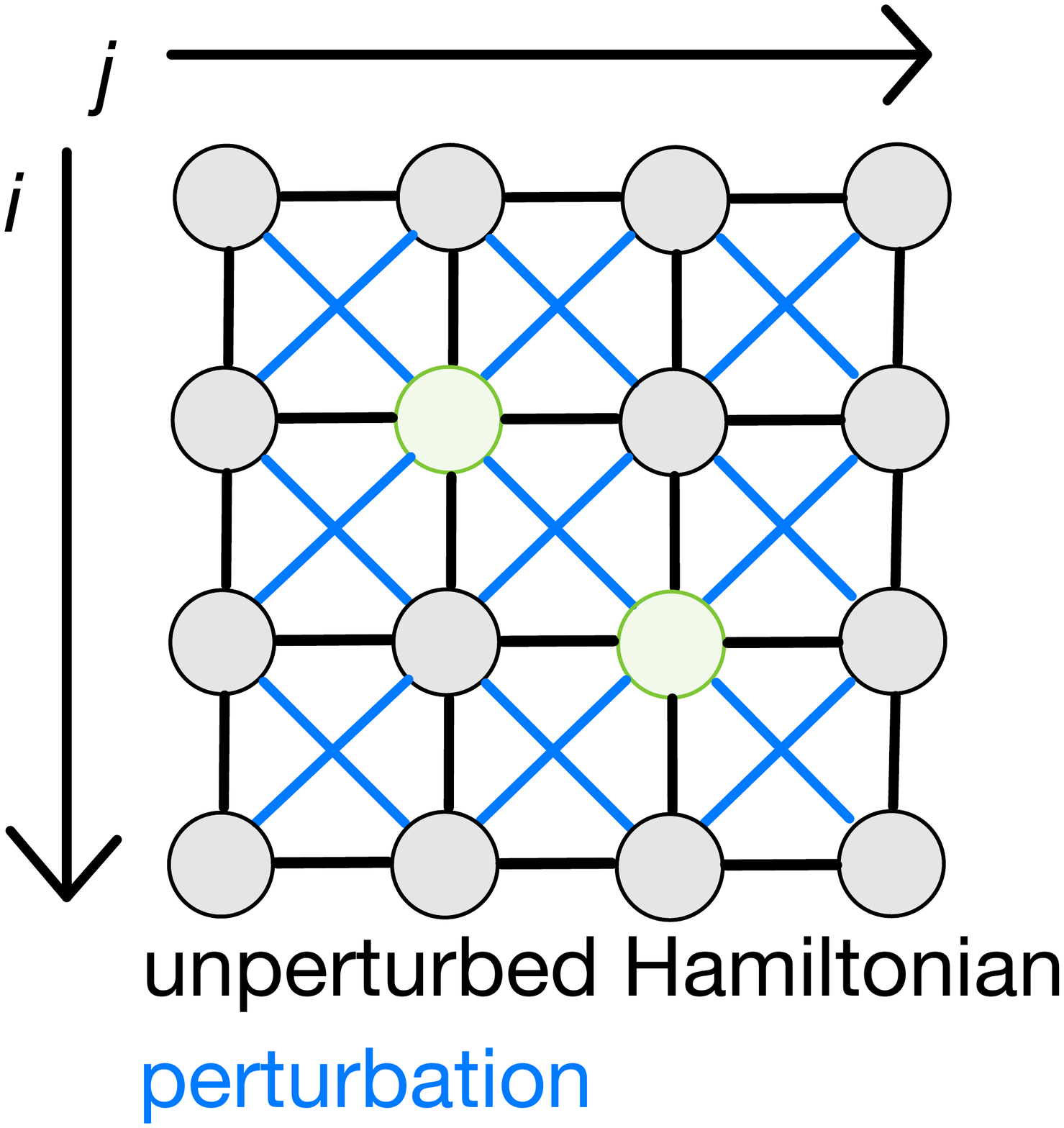}
    \caption{Sketch of the model from Section \ref{lattice}:\\ Grey circles symbolize the spin sites, the black lines represent the Heisenberg interaction of the unperturbed Hamiltonian, the blue lines represent the perturbation. The dashed lines indicate the periodic boundaries of this system. The green circles indicate the spins which are important for the observable.}
    \label{fig:skizze_RG}
\end{figure}

The last system is a spin-1/2 lattice with open boundary conditions (see Fig.\ \ref{fig:skizze_RG}).
The Hamiltonian and the perturbation are given by:
\begin{align}
    \hat{H}_0&=0.16\hat{S}^z_{1,2}+4\cdot \sum\limits_{i,j=1}^{L-1}\Vec{\hat{S}}_{i,j}\Vec{\hat{S}}_{i,j+1}+\Vec{\hat{S}}_{i,j}\Vec{\hat{S}}_{i+1,j}\label{eq:H_RG}\\
    \hat{V}&=4\sum\limits_{\alpha=x,y}\sum\limits_{i,j=1}^{L-1}\left(\hat{S}^\alpha_{i,j}\hat{S}^\alpha_{i+1,j+1} +\hat{S}^\alpha_{i+1,j}\hat{S}^\alpha_{i,j+1}\right)\label{eq:V_RG}.
\end{align}
Here, too, the original model from Ref.\ \cite{SpinGitter} is expanded by a symmetry-breaking term.\\
In this system, the correlation of the magnetization in $z$-direction of two spins is considered as an observable\
\begin{align}\label{eq:O_RG}
    \hat{O}&=4\hat{S}^z_{2,2}\hat{S}^z_{3,3}.
\end{align}
In this system we investigate the dynamics given by
\begin{align}\label{eq:LatticeRho}
    C_{\hat{O}}(t)&=\Tr{\hat{\rho}\hat{O}(t)}\\
    &\text{with}\nonumber\\
    \hat{\rho}&\propto e^{-\frac{\hat{H}^2}{2\sigma_E^2}}\mathcal{P}^+_{2,2}\mathcal{P}^+_{3,3}e^{-\frac{\hat{H}^2}{2\Delta E^2}}.
\end{align} 
The $\mathcal{P}^+_{i,j}$ operator ensures that spin on the site $i,j$ is up.\\
The second part of the equation is a Gaussian energy filter, where the standard deviation is taken from Ref.\ \cite{SpinGitter} with $\sigma_E=2$.\\
This is different from Ref.\ \cite{SpinGitter}, since there the average dynamics of an ensemble of pure states $\ket{\psi}$ was studied:
\begin{align}    
    \tilde{C}_{\hat{O}}&=\braket{\psi(t)|\hat{O}|\psi(t)}\\
    \text{with }\nonumber\\
    \ket{\psi}&\propto e^{-\frac{\hat{H}^2}{2\sigma_E^2}}\mathcal{P}^+_{2,2}\mathcal{P}^+_{3,3}\ket{\phi}\\
    \mathcal{P}^+_{i,j}&:=\hat{S}^z_{i,j}+\frac{1}{2}\hat{1},
\end{align}
where $\ket{\phi}$ is a Haar-distributed random vector. \\
The average over the dynamics of the complete ensemble is equal to the dynamics of the initial state $\hat{\rho}$.\\
The deviation of the individual dynamics from this average becomes smaller with increasing system size \cite{REIMANN2020121840,HeitmannRichterSchubertSteinigeweg+2020+421+432}.\\
This system was first listed in Ref.\ \cite{SpinGitter} as an example of a successful typicality perturbation theory for the first two approximations. In Ref.\ \cite{LongPaper}, it was then also shown that the third approximation $g_3$ is also suitable. 
The parameters of the perturbation theory for this model have already been determined in Ref.\ \cite{SpinGitter}, so that a comparison between theory and numerical results is easily accessible here.
\subsection{Conditions}
\begin{figure}[t]
    \centering
    \includegraphics[width=\linewidth]{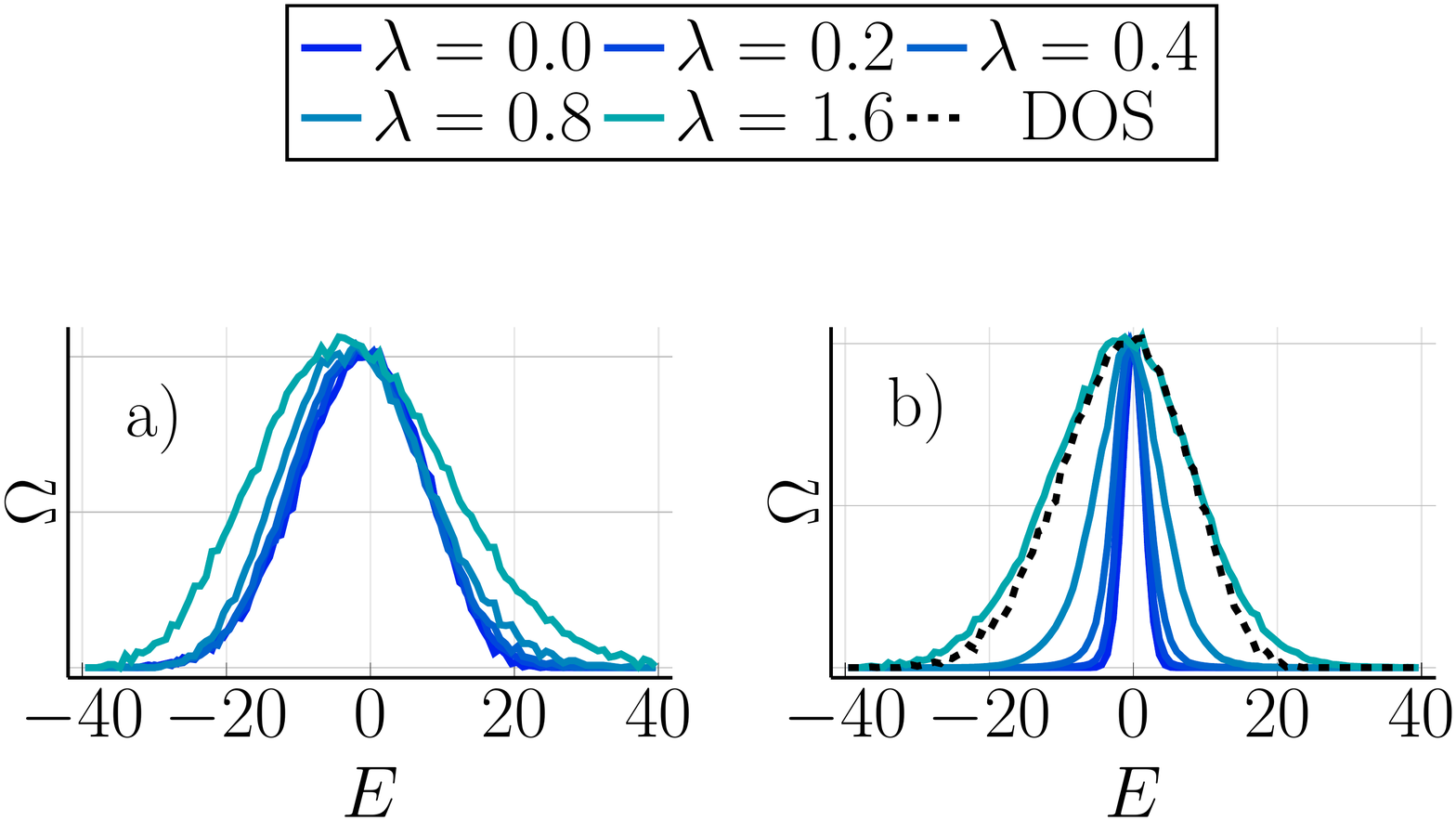}
    \caption{DOS a) or LDOS b) at different perturbation strengths $\lambda$ from the spin lattice. (Scaled for better comparability)}
    \label{fig:DOS_RG}
\end{figure}

\begin{figure}[t]
    \includegraphics[width=\linewidth]{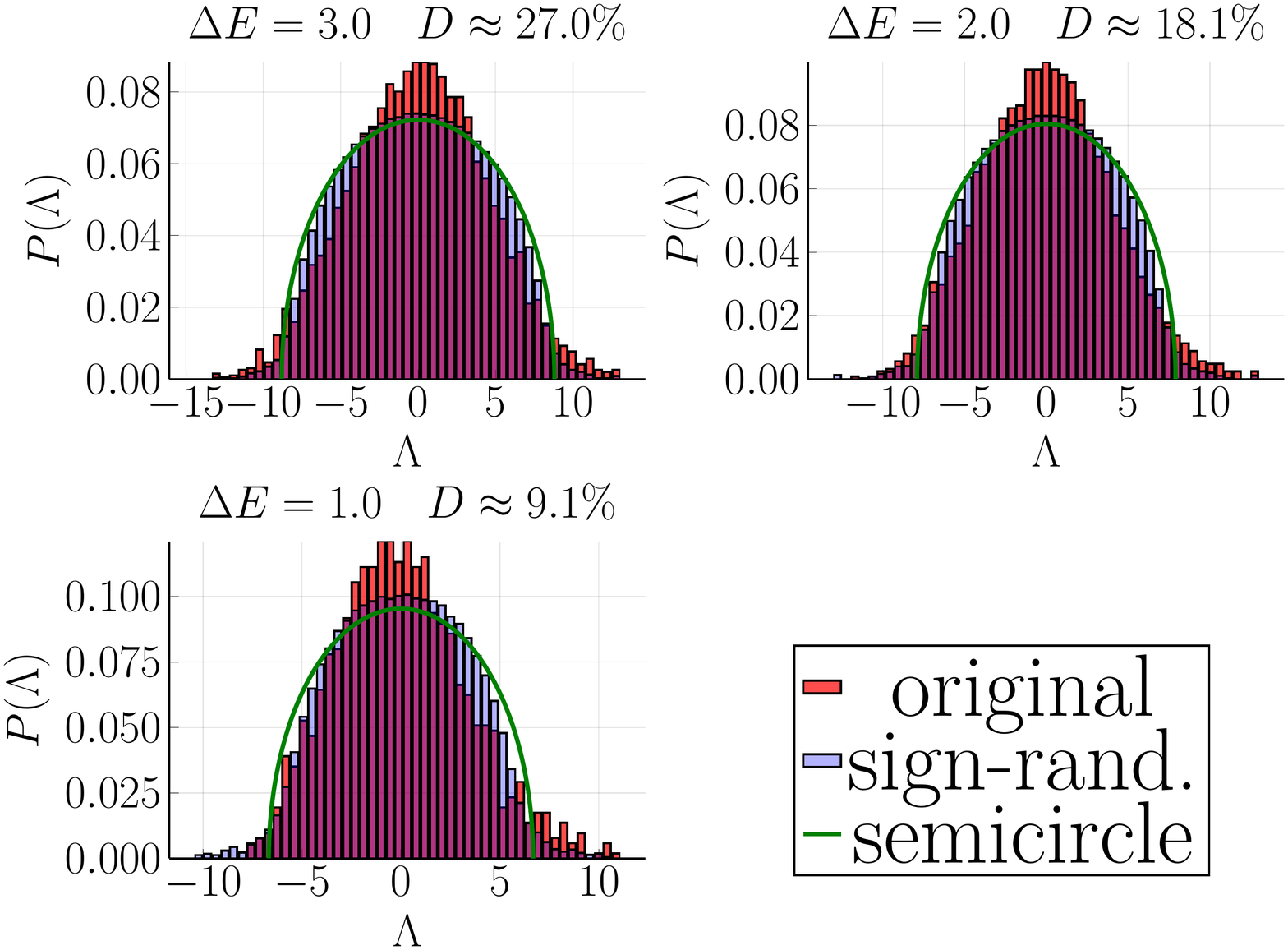}
    \caption{Spectra of the (trace free) perturbation $\hat{V}_{\Delta E}$ and the sign-randomized version $\Tilde{V}$ for different energy windows $\Delta E$. In addition the results are compared with the Wigner-semicircle law, which gives the distribution for a random matrix. $D$ indicates what percentage of the total system is in the chosen energy window.}
    \label{fig:SignRandLatt}
\end{figure}

\begin{figure}[t]
    \centering
    \includegraphics[width=\linewidth]{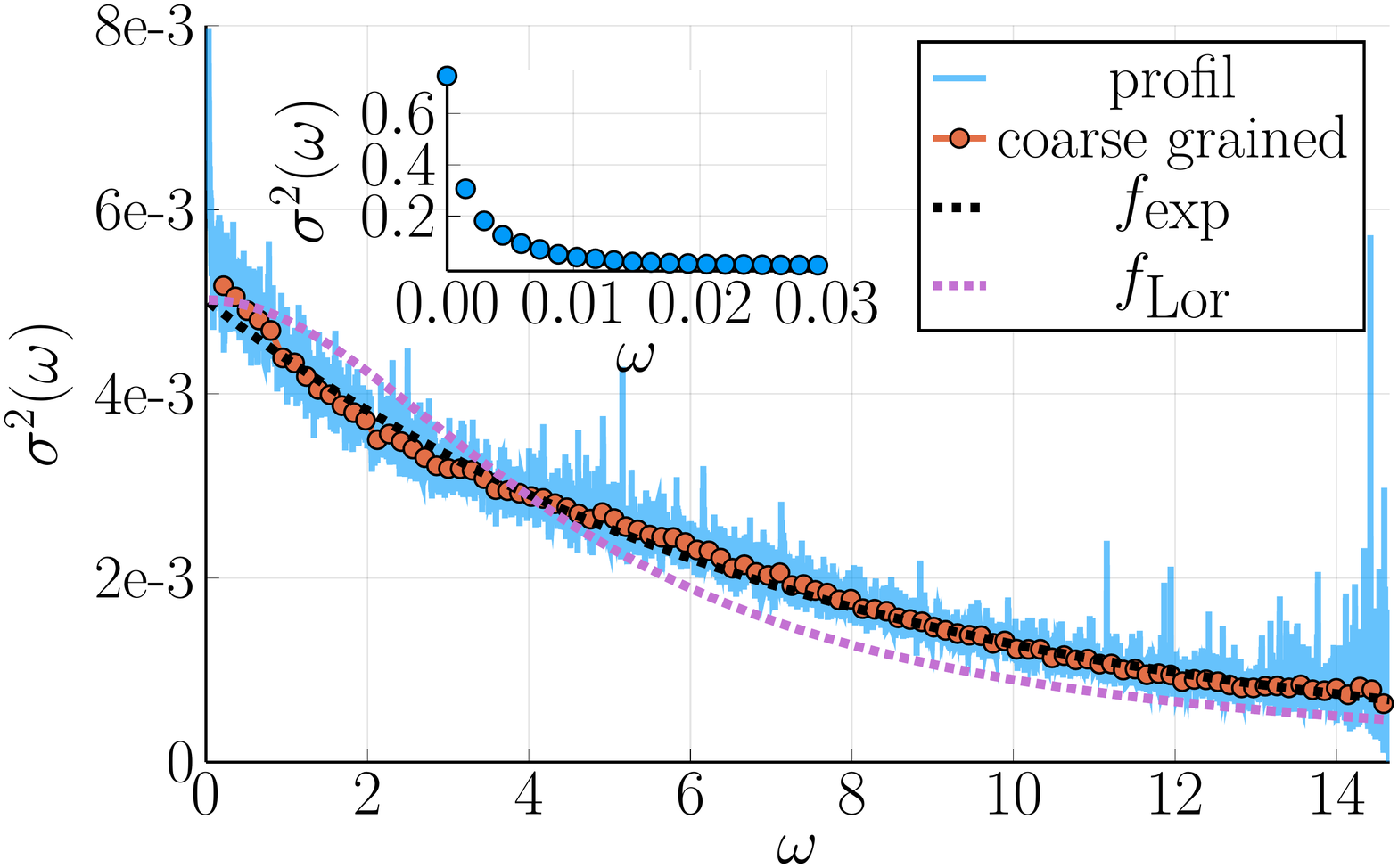}
    \caption{The perturbation-profile $\sigma^2(\omega)$ for the perturbation in Eq.\ \eqref{eq:V_RG}, with a coarse grained version. In addition, two fits can be seen, with parameters taken from Ref.\ \cite{SpinGitter}. The inset shows the variance of the low frequency elements, which are much larger than the other elements.}
    \label{fig:profil_V}
\end{figure}

Here we choose $L=4$ so that the total number of spins equals $16$. Thus, the system can also be investigated by means of ED.
In Fig.\ \ref{fig:DOS_RG} a), the DOS for various perturbation strengths $\lambda$ are shown. It can be seen that except for $\lambda=1.6$ the structure of the DOS hardly changes. Thus condition ii) is fulfilled for all weaker perturbations. \\
The LDOS can be seen in Fig.\ \ref{fig:DOS_RG} b). 
At $\lambda=1.6$, the assumption that the DOS is approximately constant failed. Even at $\lambda=0.8$ there are already non-trivial parts in the edges of the spectrum.
Thus, the condition i) is not always fulfilled.\\
To check condition iv) the sign-randomization method is used again, see Fig.\ \ref{fig:SignRandLatt}. Here we choose three different energy windows $\Delta E=\left\{3,2,1\right\}$ , such that there are windows greater, equal and smaller than the chosen standard deviation $\sigma_E$.
Like in the models before, even for the smallest energy window we see that the matrix still contains correlations.
Since the dynamics of this system was used in Ref.\ \cite{LongPaper} to test $g_3$ from Eq.\ \eqref{eq:g3}, here also the additional condition of this approximation shall be tested; namely that the perturbation profile $\sigma^2(\omega)$ is Lorentz-shaped. \\\
The dependence of $\sigma^2$ on the energy difference $\omega$ for the central $7722$ states ($60\%$ of the full spectrum) can be seen in Fig.\ \ref{fig:profil_V}. In addition, a coarse grained version and two further fitting curves are shown. The exponential curve has the form
\begin{align}
    f_{\text{exp}}(\omega)&=\sigma^2(0)\cdot e^{-\frac{\omega}{\Delta_v}},
\end{align}
where one can see, that the fitting parameter are the important parameter for $g_{l}$.\\
The fitting was performed in Ref.\ \cite{SpinGitter}, so we can use the parameters determined there:
\begin{align}
    \sigma^2(0)&=0.00502 & \Delta_v&=7.32
\end{align}
The second fitting curve is a Lorentzian, which provides the same parameters as the exponential curve. It thus has the form
\begin{align}
    f_{\textrm{Lor}}(\omega)&=\frac{\sigma^2(0)}{1+\left(\frac{\pi\omega}{2\Delta_v}\right)^2}.
\end{align}
We would like to emphasize that both fittings ignore elements close to the main diagonal (see inset of Fig.\ \ref{fig:profil_V}). This is especially worth mentioning, since $\sigma^2(0)$ is to be read near the main diagonal.\\
Besides this exception, the exponential fit describes the profile well. The Lorentzian has greater deviation from the profile than the exponential function.
\subsection{Comparison}
\begin{figure}[t]
    \includegraphics[width=\linewidth]{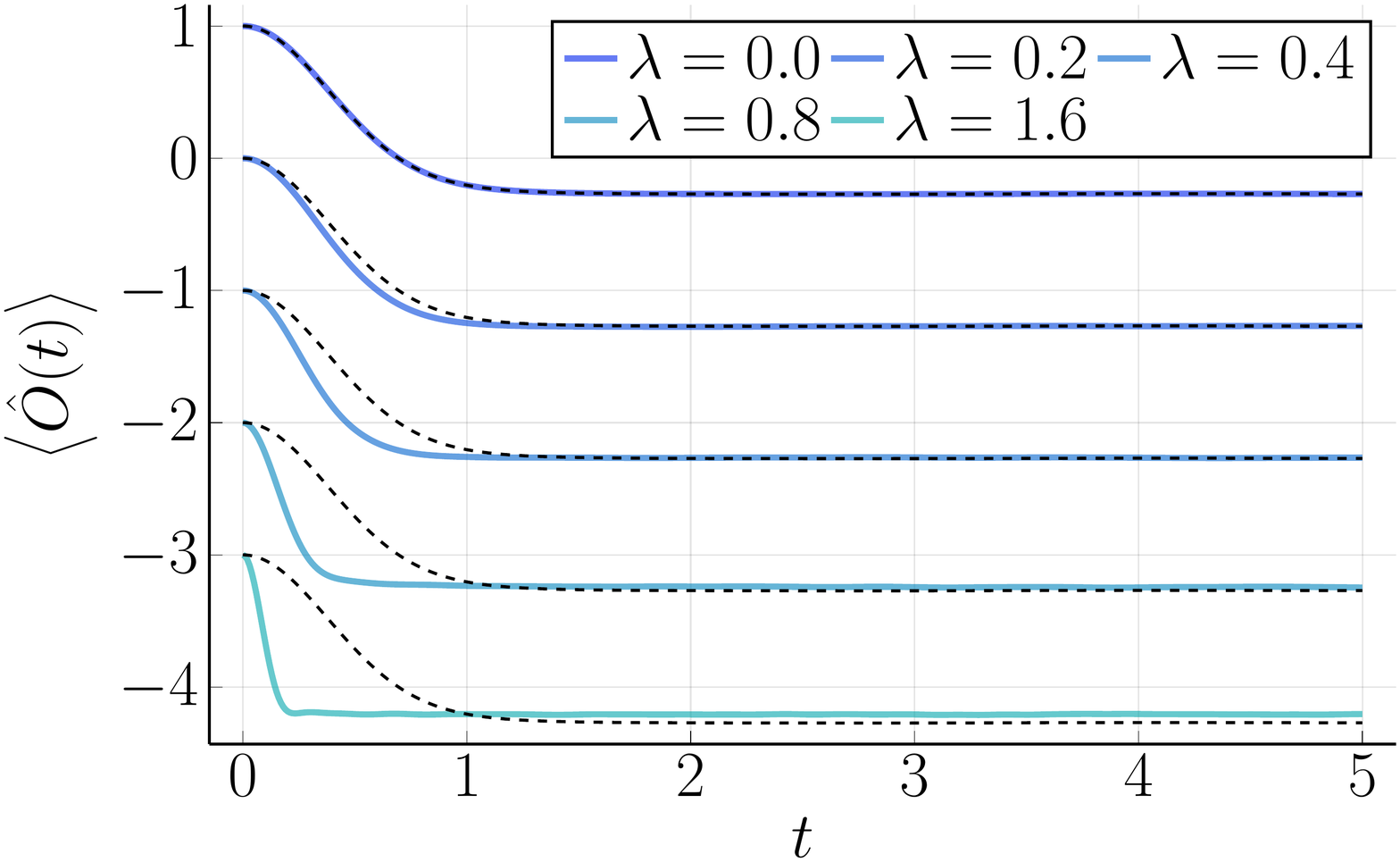} 
    \caption{Dynamics for various perturbations strength $\lambda$ of the spin lattice. The dashed lines denote the unperturbed dynamic. The curves are normalized and shifted by $1$.}
    \label{fig:Dyn_RG} 
\end{figure}

\begin{figure}
    \includegraphics[width=\linewidth]{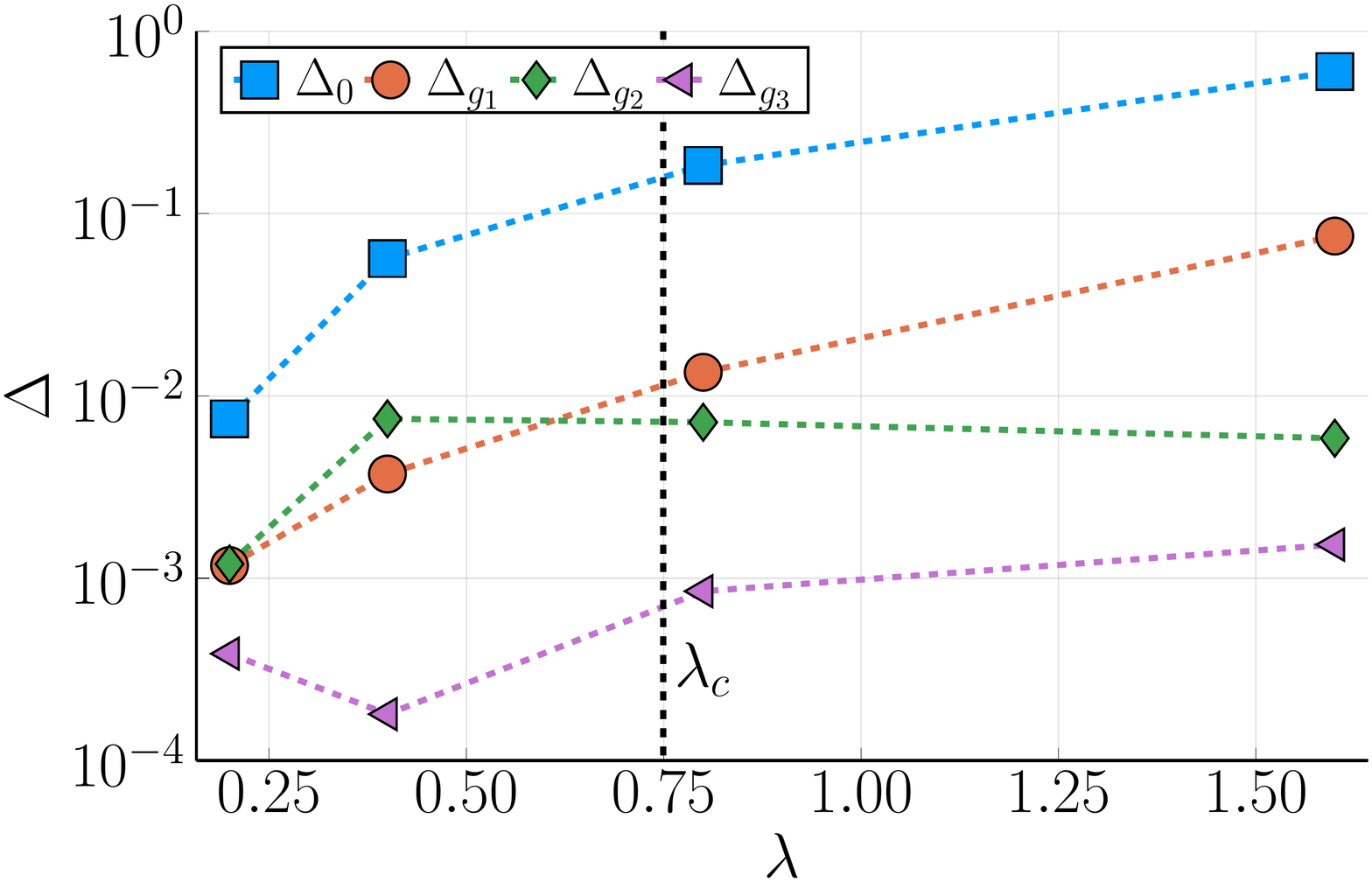} 
    \caption{Deviation between the prediction of the perturbation theory and perturbed dynamics as well as the deviation between the perturbed and unperturbed dynamics in spin lattice. $\lambda_c$ denotes the theoretical crossover from weak perturbation ($g_1,g_3$) to large perturbation ($g_2$).}
    \label{fig:Delta_RG} 
\end{figure}

In addition to the parameter of the perturbation (see \mbox{Fig.\ \ref{fig:profil_V}}), the mean level spacing
\begin{align}
\epsilon&=0.0019 
\end{align}
was also taken from Ref.\ \cite{SpinGitter}.\\
The normalized dynamics in the spin lattice can be seen in Fig.\ \ref{fig:Dyn_RG}.
In Fig.\ \ref{fig:Delta_RG} the associated deviations are plotted.\\
Since all parameters are known here, the critical perturbation strength $\lambda_c$ [cf.\ Eq.\ \eqref{eq:lambda_c}] above which $g_2$ should provide better results than $g_1$, can also be determined here. This behavior can be observed in Fig.\ \ref{fig:Delta_RG}.\\
Over all selected perturbations $g_3$ is always the best approximation. Moreover, the expectations on $g_1$ and $g_2$ are fulfilled: $g_1$ becomes more accurate the weaker the perturbation becomes, while $g_2$ provides the dynamics at large perturbations.
Since in these models, unlike the first two, the parameters were determined exactly and were not treated as fitting parameters, the fact that $g_1$ works better than $g_2$ in the first two models for larger perturbations could also be an artifact of the fitting process.
\subsection{Conclusion}
In the last system, presented in Refs.\ \cite{SpinGitter,LongPaper} as an example of the effectiveness of perturbation theory, \mbox{condition i)} is fulfilled by the use of energy filters for perturbation strengths up to $\lambda=0.4$.
Condition ii) also seems to be fulfilled for higher perturbation strengths up to $\lambda=0.8$.
As in the other two systems, correlations can be detected even within small energy windows, therefore condition iv) is not fulfilled.\\
An analysis of the profile of the perturbation shows that it approximately represents an exponential function and not a Lorentzian, as assumed in the derivation of Eq.\ \eqref{eq:g3}. This can be interpreted either as a sign of the robustness of the theory (as in Ref.\ \cite{LongPaper}), or as a sign that the theory is merely working randomly here. \\
The perturbation theory gives good results in this system even without free fitting parameters.
\section{Comparison of the energy window with an mesoscopic case}\label{mesos}
Since the size of the energy window (or strength of the filter), seems to have a influence on the accuracy of the perturbation theory (see Section \ref{crossladder}), the question arises what window size would be realistic for mesoscopic systems.
A simple example is used to estimate orders of magnitude:\\
Consider two blocks of iron with a mass of $0.5\si{\gram}$ each and each block has a temperature which deviates from the mean temperature $T=273.15\si{\kelvin}$ at the beginning. Upon contact with each other, the temperature of both blocks relaxes close to the mean temperature within seconds. Using standard textbook methods \cite{Fliebach2018}, it can be shown that the variance of the energy in the iron block is $\sigma_E^2=k_BC_vT^2$.\\\
This results in a range of relevant frequencies of \mbox{$\omega_\textrm{rel.}=\sqrt{k_BC_v}\frac{2\cdot T}{\hbar}$}, where $C_v\approx0.45\si{\joule\per\kelvin}$ is the heat capacity of both blocks together \cite{Binder1999-us},$\hbar$ is the reduced Planck constant and $k_B$ is the Boltzmann constant.
From these values it follows $\omega_\textrm{rel.}\approx1.30\cdot10^{25}\si{\per\second}$. If we compare this with the typical time scale of such a system (e.g. the relaxation time $\tau$), it is evident that
\begin{align}
    \omega_\textrm{rel.}\cdot\tau\gg1.
\end{align}
For the first system the relaxation time is $\tau=26.2$ in the case of no perturbation and energy cutting
(see \mbox{Section \ref{crossladder}}). So the pro\-duct of this time and the chosen energy windows (which equals ${\omega_\textrm{rel.}}/{2}$) results in
\begin{align}
    \Delta E\cdot \tau&=26.7\cdot\left\{\frac{\pi}{2},\frac{\pi}{5},\frac{\pi}{10}\right\}\\
    &\approx\left\{41.94,16.78,8.39\right\}.
\end{align} 
It can be seen that these products are considerably smaller than in the mesoscopic case. The energy window with the best results for the perturbation theory ($\Delta E=\frac{\pi}{10}$), is the furthest from the mesoscopic case. It is therefore questionable whether this case is relevant for everyday situations.\\
For the filtered lattice (Section \ref{lattice}), it can be seen that $\tau\cdot\sigma_E=2.68$. Thus, this system is even further away from the mesoscopic estimation.\\
While the model in Section \ref{chainladder} has no filter nor energy window, we use the standard deviation of the full (unperturbed) Hamiltonian $\sigma_{\hat{H}}$ to specify the range of relevant frequencies:
\begin{align}
    \omega_{rel.} &=2\sigma_{\hat{H}}=2\cdot2.24
\end{align}However, this system has not only one typical time scale, but two:\\
According to our definition of the relaxation time we get $\tau=186.4$ [see below Eq.\ \ref{eq:Delta}]. However, in Fig.\ \ref{fig:Dyn_JS} it is easy to see a large part of the dynamics has already happened before $t=30$.
This phenomenon is called prethermalization (see \cite{Mori_2018} for a summary) and can appear in systems close to integrability. This is the case in this system; the symmetry-breaking term $h$ breaks integrability [cf.\ Eq.\ \eqref{eq:h}].
Thus, two estimates are possible here, once using the reflexation time as the time scale and then using the prethermalization time ($\tau_{\text{pre.}}\approx 30$):\\
\begin{align}
    \omega_{\text{rel.}}\cdot \tau&=2\cdot2.242\cdot186.4\\
    &\approx835.0\\
    \omega_{\text{rel.}}\cdot\tau_{\text{pre.}}&=2\cdot2.242\cdot30\\
    &\approx134.4
\end{align}
In both cases, the result is closer to the mesoscopic estimate than the other two systems.
\section{Final Conclusion}\label{results}
In general, no system could be found that fulfills all conditions. For example, correlations could be demonstrated for all models. \\
However, the effects of those correlations are difficult to assess. \\
The first model shows that despite the existing correlations, the perturbation theory provides an accurate description for small windows. However, the behavior can be understood by softening the window (see the end of Section \ref{crossladder}). Moreover, it is questionable whether this window size is relevant at all for mesoscopic scaling. \\
The second model shows good agreement with the perturbation theory despite correlations and non-constant DOS. However, good results were also achieved here with rather arbitrary choices for $g$ without a theoretical basis.\\
In the last model the parameters are known in contrast to the first two models, where the parameters were treated as fitting parameters.
This model shows that those parameters achieve good results in the perturbation theory framework, despite existing correlations. In particular, it could be shown that the perturbation-profile $\sigma^2(\omega)$ does not strictly fulfill all conditions for the application of $g_3$, but nevertheless provides a good description of the perturbed dynamics. However, also in this case it is questionable how relevant the chosen energy window is.\\
An investigation of finite size effects would be interesting, but is outside the scope of this work.\\
In summary, how the violation (or fulfillment) of the conditions inflict the validness of the perturbation theory is unclear. Only in one case a small energy window seems to improve the result. But even in this case, the improvement can partly be explained without the perturbation theory.\\
\section{acknowledgement}
We thank R. Heveling for helpful comments on the manuscript. This work was supported by the Deutsche Forschungsgemeinschaft (DFG) within the Research Unit FOR 2692 under Grant No.\ 397107022 (GE 1657/3-2).
% and No.\ 397067869 (STE 2243/3-2).
\bibliography{bib.bib}
\end{document}